\documentclass[pra, twocolumn, superscriptaddress, longbibliography, floatfix]{revtex4-1}
\usepackage{amsmath, amsfonts, amssymb, mathrsfs}
\usepackage{graphicx, subfigure, color}
\usepackage{bm}
\usepackage{braket}
\usepackage{epstopdf}
\usepackage[normalem]{ulem} % needed for \scrap-command
\usepackage{color}
\usepackage[breaklinks]{hyperref}
\usepackage{lipsum}
\usepackage{xcolor}
\usepackage{enumerate}

\definecolor{grey}{rgb}{.6,.6,.6}

\hypersetup{
   colorlinks,
   citecolor=blue,
   filecolor=blue,
   linkcolor=blue,
   urlcolor=blue
}

\newcommand\rev[1]{{\color{black}#1}}

\begin{document}

\title{Anyonic tight-binding models of parafermions and of fractionalized fermions}

\author{Davide Rossini}
\affiliation{Dipartimento di Fisica dell'Universit\`a di Pisa and INFN, Largo Pontecorvo 3, I-56127 Pisa, Italy}

\author{Matteo Carrega}
\affiliation{NEST, Istituto Nanoscienze-CNR and Scuola Normale Superiore, Piazza San Silvestro 12, 56127 Pisa, Italy}

\author{Marcello Calvanese Strinati}
\affiliation{Department of Physics, Bar-Ilan University, 52900 Ramat-Gan, Israel}

\author{Leonardo Mazza}
\affiliation{LPTMS, UMR 8626, CNRS, Universit\'e Paris-Sud, Universit\'e Paris-Saclay, 91405 Orsay, France}

\begin{abstract}
  Parafermions are emergent quasi-particles which generalize Majorana fermions and possess intriguing anyonic properties.
  The theoretical investigation of effective models hosting them is gaining considerable importance
  in view of present-day condensed-matter realizations where they have been predicted to appear. 
  Here we study the simplest number-conserving model of particle-like Fock parafermions, namely a one-dimensional
  tight-binding model. By means of numerical simulations based on exact diagonalization and on the density-matrix
  renormalization group, we prove that this quadratic model is nonintegrable and displays bound states in the spectrum,
  due to its peculiar anyonic properties. Moreover, we discuss its many-body physics, characterizing anyonic correlation
  functions and discussing the underlying Luttinger-liquid theory at low energies. In the case when Fock parafermions
  behave as fractionalized fermions, we are able to unveil interesting similarities with two counter-propagating edge modes
  of two neighboring Laughlin states at filling $1/3$. 
\end{abstract}

\date{\today}
\maketitle

\section{Introduction}

Anyons, namely emergent quasi-particles with a quantum statistics that is neither bosonic nor fermionic,
are one of the most fascinating concepts in condensed-matter physics~\cite{Leinaas_1977, Wilczek_1982}.
They are the hallmark of non-trivial topological phases of matter emerging in strongly-correlated two-dimensional systems,
the most famous example being the fractional quantum Hall
effect~\cite{Wen_1995, Tsui_1982, Stern_2008, Dolev_2010, Carrega_2011}.
In such case, anyons appear in the gapped bulk of a system with non-zero topological order and,
thanks to the so-called bulk-boundary correspondence~\cite{Wen_1995, Wilczek_1982, Blasi_2012},
are responsible for the chiral metallic edge states, featuring peculiar transport properties.
Different topological phases have been predicted~\cite{Wen_1995, Stern_2008}, depending on the value of the filling
factor $\nu$, showing that anyons can support fractional charges and fractional statistics. 
A prominent role has been played by the $\nu=5/2$ case~\cite{Stern_2008, Dolev_2010, Carrega_2011},
whose low-energy quasi-particles are believed to possess non-Abelian
statistics~\cite{Wilczek_1982, Moore_1991, Levin_2007},
supporting chiral Majorana fermions on the edge. A plethora of other platforms, where non-trivial topological phases exist,
have been recently put forward, thus triggering a new field of investigation in theoretical
as well as experimental condensed-matter physics.

One-dimensional (1D) anyonic models have been object of extensive theoretical studies in the last
decades~\cite{Bogoliubov_1992, Amico_1998, Kundu_1999, Batchelor_2006, Calabrese_2007, Santachiara_2007, Pellegrino_2007, Patu_2007, Feiguin_2007, Santachiara_2008, RenGui_2008, Bellazzini_2009, Keilmann_2011, Santos_2012, Greschner_2015, Tang_2015, Hao_2016, Marmorini_2016, Arcila-Forero_2016, Colcelli_2018}.
In this context, generalizations of Majorana fermions, dubbed parafermions or fractionalized Majorana fermions,
have been introduced~\cite{Alicea_2016}. They possess a fractional anyonic statistics which can be exploited for performing topological
quantum computation, thus enhancing their potentialities, with respect to those of Majorana fermions~\cite{Nayak_2008, Clarke_2013}.
Moreover, they have been predicted to form in some hybrid systems, thanks to the interplay between superconductivity
and other strongly correlated systems~\cite{You_2012, Lindner_2012, Cheng_2012, Clarke_2013, Vaezi_2013, Barkeshli_2013, Klinovaja_2014, Barkeshli_2014a, Thakurathi_2017, VinklerAviv_2017, Santos_2017, Alavirad_2017, Wu_2018, Mazza_2018b, Calzona_2018, Fleckenstein_2018, Guiducci_2018}.
The anyonic statistics of parafermions is encoded in operators whose commutation relations are governed
by the presence of an angle $0\leq \kappa\leq 1$. Such kind of operators can be employed to describe
$\mathbb Z_p$-symmetric models with exotic critical properties~\cite{Albertini-Kedem, Mong_2014, Mazza_2018, Samajdar_2018},
as well as topological models with zero-energy boundary modes~\cite{Fendley_2012, Burrello_2013, Motruk_2013, Bondesan_2013, Zhuang_2015, Alexandradinata_2016, Sreejith_2016, Xu_2017, Moran_2017, Munk_2018}.

The formalism of Fock parafermions (FP) allows for the discussion of parafermions using a simple and intuitive
particle-like picture~\cite{Cobanera_2014}, that has been already exploited in the study of topological
and non-topological parafermionic zero-energy modes~\cite{Cobanera_2015, Iemini_2017, Chew_2018, Calzona_2018}.
FPs are generically labeled by a natural number $p\geq2$ (for $p=2$, they are canonical complex fermions),
which determines their statistical parameter $\kappa = 2/p$~\cite{Cobanera_2014}.
For even values of $p$, clusters of FPs behave exactly as fermions~\cite{Cobanera_2017}, so that FPs can be interpreted
as fractionalized fermions.

The main motivation of this article is to understand whether simple lattice Hamiltonians of FPs can be employed
to model possible condensed-matter setups displaying non-trivial topological order. 
To this purpose, we explore a basic number-conserving tight-binding chain of FPs
(notice that previously considered FP models do not conserve the particle number).
Despite its formal simplicity, the anyonic statistical properties of FPs make this quadratic model nonintegrable,
and thus not amenable to exact analytic treatments.
By means of a density-matrix renormalization group (DMRG) based analysis~\cite{Schollwoeck_2005}, we show that
several distinguishing features of these physical objects can be spotlighted, giving new hints on the nature of FPs. 
For instance, explicitly neglecting interactions (namely, quartic terms in the Hamiltonian) permits to stress the role of anyonic statistics. 
We also compare the properties of our model with those of known anyonic models, in order to underline its peculiarities.

Specifically, we are going to focus on $p=3$, a value that yields the simplest non-trivial model of parafermions, and on $p=6$.
The latter value, being even, allows for the definition of genuine fermionic observables (emerging from clustering three FPs).
In such case we show that fermionic observables display properties that cannot be easily
traced back to a simple fermionic model~\cite{Chang_2003}.
By comparing our numerics with the prediction of the hydrodynamic theory for the boundary of fractional
quantum Hall states~\cite{Wen_1990a, Wen_1995}, we unveil strong analogies between our model
and two counter-propagating edge modes of neighboring Laughlin states at filling $1/3$.
The importance of this latter setup in the development of schemes to localize zero-energy parafermionic
modes~\cite{Lindner_2012, Cheng_2012, Clarke_2013} paves the way to further applications of our tight-binding model.

This article is organized as follows.
In Sec.~\ref{Sec:Model} we briefly recall the formalism of FPs, introduce our tight binding model of FPs,
and discuss its relation with other anyonic models.
Before discussing the main results, we present an analysis of the 
one- and two-body physics as a gentle introduction to the many-body case (Sec.~\ref{Sec:FewBody}),
and demonstrate the nonintegrability of the model through its level spacing statistics (LSS) (Sec.~\ref{Sec:LSS}).
The bulk of the paper is constituted by Sec.~\ref{Sec:Z3}, where we show the results of DMRG simulations for the many-body problem,
with emphasis on the anyonic correlation functions. The potential relevance of our model in describing, on a lattice,
the boundary between two neighboring quantum Hall bars is discussed relying on a phenomenological low-energy approach.
Finally, Sec.~\ref{Sec:Conc} is devoted to the conclusions.

\section{Model}\label{Sec:Model}

\subsection{Fock parafermions}
\label{Sec:FP}

We consider a set of $2L$ parafermions $\{ \hat \gamma_{j}\}$ of order $p$ ($p\in \mathbb N$ and $p\geq 2$),
satisfying the following algebra:
\begin{equation}
  \hat \gamma_j \hat \gamma_l = \omega^{\text{sgn}(j-l)} \hat \gamma_l \hat \gamma_j,
  \quad \mbox{with } \;\; \omega = e^{2\pi i /p},
\end{equation}
and also
\begin{equation}
\hat \gamma_j^p = 1,
  \qquad 
  \hat \gamma_j^\dagger = \hat \gamma_j^{p-1}.
\end{equation}
For $p=2$, the $\{ \hat \gamma_j \}$ are a set of Majorana modes obeying a Clifford algebra.
Since fermionic systems can be equivalently described using the complex-fermion representation
$\hat c_j^{(\dagger)} = \tfrac12 (\hat \gamma_{2j-1} \pm i \hat \gamma_{2j})$, the authors of Ref.~\cite{Cobanera_2014}
have introduced FP operators $\hat F_j^{(\dagger)} $, which allow for an analogous particle-like description of parafermionic systems. 
For $p>2$ the transformation becomes non-linear and reads:
\begin{equation}
  \hat F_j = \frac{p-1}{p} \hat \gamma_{2j-1} - \frac{1}{p} \sum_{m=1}^{p-1} 
  \omega^{m (m+1)/2}\,\rev{{(i)}^m} \, \hat \gamma_{2j-1}^{m+1} \, \hat \gamma_{2j}^{\dagger m} .
\end{equation}

If one considers a single site, a local Fock space of dimension $p$ is associated to each pair
of operators $\hat F_j^{(\dagger)}$, with basis states
\begin{equation}
  \ket{m_j} = \hat F^{\dagger m}_j\ket{0}, \qquad 0 \leq m \leq p-1.
\end{equation}
Here $\ket {m_j}$ labels the state with $m$ parafermions on site $j$, and indeed it is an eigenstate
of the density operator
\begin{equation}
  \hat N_j = \sum_{l=1}^{p-1} \hat F^{\dagger l}_j \, \hat F^l_j
  \label{eq:Nop}
\end{equation}
with eigenvalue $m$.
Thus, on each site, the system can accommodate up to $p-1$ parafermions.
The operators $\hat F_j$ and $\hat N_j$ have the following representations in the Fock basis $\{ \ket{m_j}\}_{m=0}^{p-1}$:
\begin{equation}
  \hat F_j = \left(\begin{matrix}
    0 & 1 & 0 & \cdots & 0 \\
    0 & 0 & 1 & \cdots & 0 \\
    0 & 0 & 0 & \cdots & 0 \\
    \vdots & \vdots & \vdots & & \vdots \\
    0 & 0 & 0 & \cdots & 1 \\
    0 & 0 & 0 & \cdots & 0 
  \end{matrix} \right),
  \quad 
  \hat N_j = \left(\begin{matrix}
    0 & 0 & 0 & \cdots & 0 \\
    0 & 1 & 0 & \cdots & 0 \\
    0 & 0 & 2 & \cdots & 0 \\
    \vdots & \vdots & \vdots & & \vdots \\
    0 & 0 & 0 & \cdots & 0 \\
    0 & 0 & 0 & \cdots & p-1 
  \end{matrix} \right).
  \label{Eq:Matrix:Elements}
\end{equation}
Mathematically, they obey the following relations (among several others):
\begin{equation}
  \hat F^p_j = 0, \qquad
  \hat F^{\dagger m}_j \hat F_j^m + \hat F_j^{p-m} \hat F_j^{\dagger (p-m)} = 1.
  \label{Eq:FP:CR1}
\end{equation}

Considering different sites, FP operators obey anyonic commutation relations:
\begin{equation}
  \hat F_j \hat F_l = \omega^{\text{sgn}(l-j)} \hat F_l \hat F_j, \qquad 
  \hat F_j^\dagger \hat F_l = \omega^{- \text{sgn}(l-j)} \hat F_l \hat F_j^\dagger,
  \label{Eq:FP:CR2}
\end{equation}
and the statistical parameter $\kappa$, defined by rewriting the previous relation
as $\hat F_j \hat F_l = e^{i \pi \kappa \, \text{sgn}(l-j)} \hat F_l \hat F_j$, is $\kappa = 2/p$.
The full Hilbert space has dimension $p^L$,
being the tensor product of the Fock spaces associated to each site.

One of the interesting properties of FPs is that in some cases they can be considered as roots
of fermionic operators~\cite{Cobanera_2017}. Indeed, when $p = 2m$, the operator $\hat f_j = \hat F_j^m$ satisfies
canonical anticommutation relations:
\begin{equation}
  \{\hat f_j,\, \hat f_l \} = 0, \qquad
  \{\hat f_j,\, \hat f_l^\dagger \}= \delta_{j,l},  \qquad
  \hat f_j^2 = 0.
 \label{Eq:Fermi:p6}
\end{equation}
As such, FP models offer the unique possibility of studying genuine fermionic observables in lattice models of fractionalized fermions.

\subsection{The Hamiltonian and its symmetries}
\label{Sec:Ham:Sym}

In this paper we focus on a 1D tight-binding model of FPs, described by the Hamiltonian
\begin{equation}
  \hat H = -t \sum_j \left[ \hat F_j^\dagger \hat F_{j+1} + \hat F_{j+1}^\dagger \hat F_{j} \right],  \quad t>0.
  \label{Eq:Ham}
\end{equation}
The parameter $t$ can be fixed to one, thus setting the system's energy scale.
The model enjoys a U(1) symmetry related to the conservation of the total number of particles $\hat N = \sum_j \hat N_j$.
It is not inversion invariant, because of the asymmetric commutation relations~\eqref{Eq:FP:CR2},
so that $\hat F_j \to \hat F_{-j}$ is not a canonical transformation that preserves the algebra of FPs.
Moreover it is not time-reversal invariant, because the anyonic statistics of FPs breaks time-reversal invariance
(indeed, applying a Fradkin-Kadanoff transformation~\cite{Fradkin_1980}, the model does not enjoy a matrix representation
with real entries).
Finally, we observe that it is not particle-hole symmetric, because $\hat F_j \to \hat F^\dagger_j$
is a transformation that does not conserve the parafermionic algebra. Yet, it enjoys a symmetry which is the combination
of particle-hole and inversion symmetry: $\hat F_j \to \hat F^\dagger_{-j}$.
For this reason, it is possible to confine the analysis to densities $N/L \leq (p-1)/2$.

\subsection{Comparison with previously considered anyonic models}\label{Sec:Relation}

\textit{Integrability.}---
Differently from most of the 1D anyonic models studied so far (see, e.g.,
Refs.~\cite{Amico_1998, Calabrese_2007, Santachiara_2007, Patu_2007, Feiguin_2007}), Hamiltonian~\eqref{Eq:Ham}
can be shown to be nonintegrable. Indeed, parafermionic operators written in momentum space do not satisfy an easy algebra
(this significantly contrasts with the cases of bosons and fermions). However, there exist anyonic models which enjoy
exact solvability through Bethe ansatz. In Sec.~\ref{Sec:LSS}, we rule out this possibility by looking at the LSS
of the Hamiltonian spectrum.

\textit{Relation with fermionic and bosonic models.}---
An important part of the literature deals with anyons that are obtained by modifying bosonic models,
as for the anyonic Lieb-Liniger~\cite{Calabrese_2007} or the anyon-Hubbard model~\cite{Keilmann_2011}.
There, for $\kappa = 0$ the model is bosonic, but for $\kappa = 1$ it is not fermionic, although the operators
anticommute (an exception is the case in which infinitely-repulsive on-site interactions are considered,
where a fermionic limit can be identified). 
In our case, for $\kappa = 1$ ($p=2$) an exact fermionic limit is recovered. However, for $\kappa \to 0$ ($p \to \infty$),
the model is not bosonic, although the operators commute. The reason lies in the precise matrix elements
of the operator $\hat F_j$ displayed in Eq.~\eqref{Eq:Matrix:Elements}, that do not possess the proper bosonic enhancement.
\rev{Indeed, an ordinary bosonic annihilation operator $\hat b_j$ obeys the following relation:
  $\hat b_j |n_j\rangle = \sqrt{n_j}\,|n_j-1\rangle$.
  Therefore its matrix representation in the Fock basis would have entries
  $\{ 1, \sqrt{2}, \ldots, \sqrt{p-1} \}$ along its first upper diagonal,
   instead of a list of ones.}
  
\textit{Fractionalization.}---
As we already stressed, in a model of FPs with even $p$, it is possible to study the behavior of well-defined
fermionic observables~\cite{Cobanera_2017}. To our knowledge, this is a unique feature of our anyonic model.

\section{One- and two-body physics}\label{Sec:FewBody}

Before analyzing the actual many-body properties of Hamiltonian~\eqref{Eq:Ham},
we focus on its one- and two-body physics.
When just a single parafermion is considered ($N=1$), the statistics is irrelevant
and thus the model trivially reduces to a nearest-neighbor hopping of one particle in a 1D lattice.
The system can be directly diagonalized after defining the momentum-space operators
\begin{equation}
  \hat F_{k} = \frac{1}{\sqrt{L}} \sum_j e^{i k j} \hat F_j.
  \label{eq:Fourier}
\end{equation}
Indeed, the eigenstates of Eq.~\eqref{Eq:Ham} in the subspace with one particle are
\begin{equation}
  \ket{k} = \hat F^\dagger_k \ket{0},
\end{equation}
where $\ket{0}$ denotes the vacuum state, and are associated to the eigenvalues
\begin{equation}
  \varepsilon(k) = -2 t \cos(k),
  \qquad k = \frac{2 \pi m}{L} , \; (m \in \mathbb Z_L) .
\end{equation}
The quantization of momenta follows by imposing periodic boundary conditions (PBC).

Conversely, if one considers a higher number of parafermions, their anyonic statistics becomes important.
We remark that the $\hat F_k$ operators in momentum space are not FP operators, since they obey
an algebra which is different from the relations in~\eqref{Eq:FP:CR1} and~\eqref{Eq:FP:CR2}.
Therefore, even if at a formal level, the Hamiltonian~\eqref{Eq:Ham} can be rewritten in
a diagonal form as $\hat H = \sum_k \varepsilon(k) \hat F_k^\dagger \hat F_k$,
the model cannot be easily solved, because it is not the sum of commuting terms,
namely
\begin{equation}
  [\hat F_k^\dagger \hat F_k, \hat F_q^\dagger \hat F_q] \neq C_k \delta_{k,q}.
\end{equation}
Let us then proceed by steps and solve the model in the two-particle sector ($N=2$),
for which the most generic form of the wavefunction reads:
\begin{equation}
  \label{Eq:2bodyAnsatz}
  \ket {\Psi} = \sum_{1 \leq n_1 \leq n_2 \leq L} a(n_1, n_2) \ket{n_1, n_2},
\end{equation}
where $\ket{n_1, n_2} = \hat F_{n_1}^\dagger \hat F_{n_2}^\dagger \ket{0}$ (with $n_1 \leq n_2$).

%%%%%%%%%%%%%%%%%%%%%%%%%%%%%%%%%%%%%%%%%%%%%%%%%%%%%%%%%%%%%%%%%%%%%%%%%%%%%%%%
\begin{figure}[!t]
  \includegraphics[width=\columnwidth]{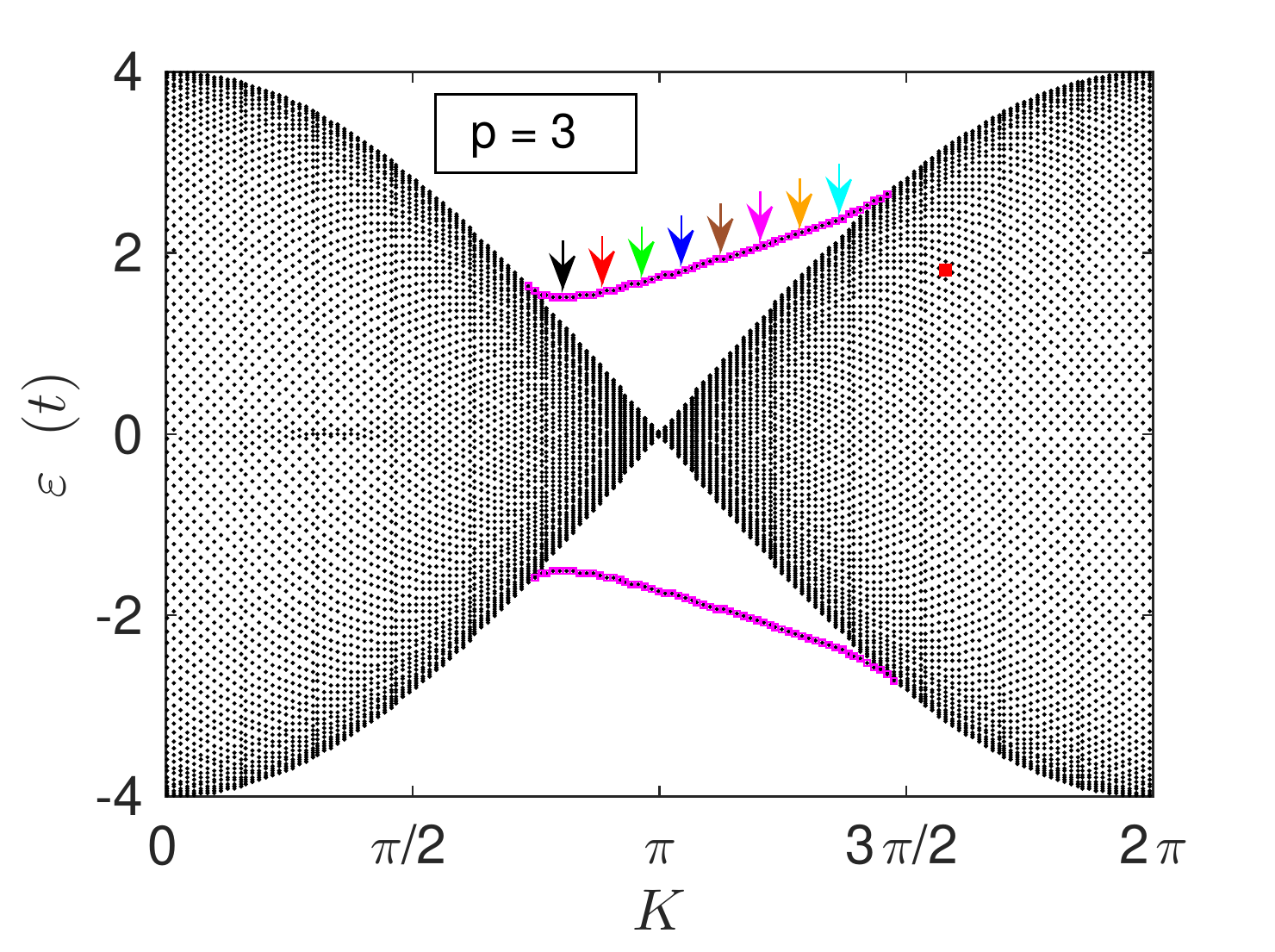}
    \includegraphics[width=\columnwidth]{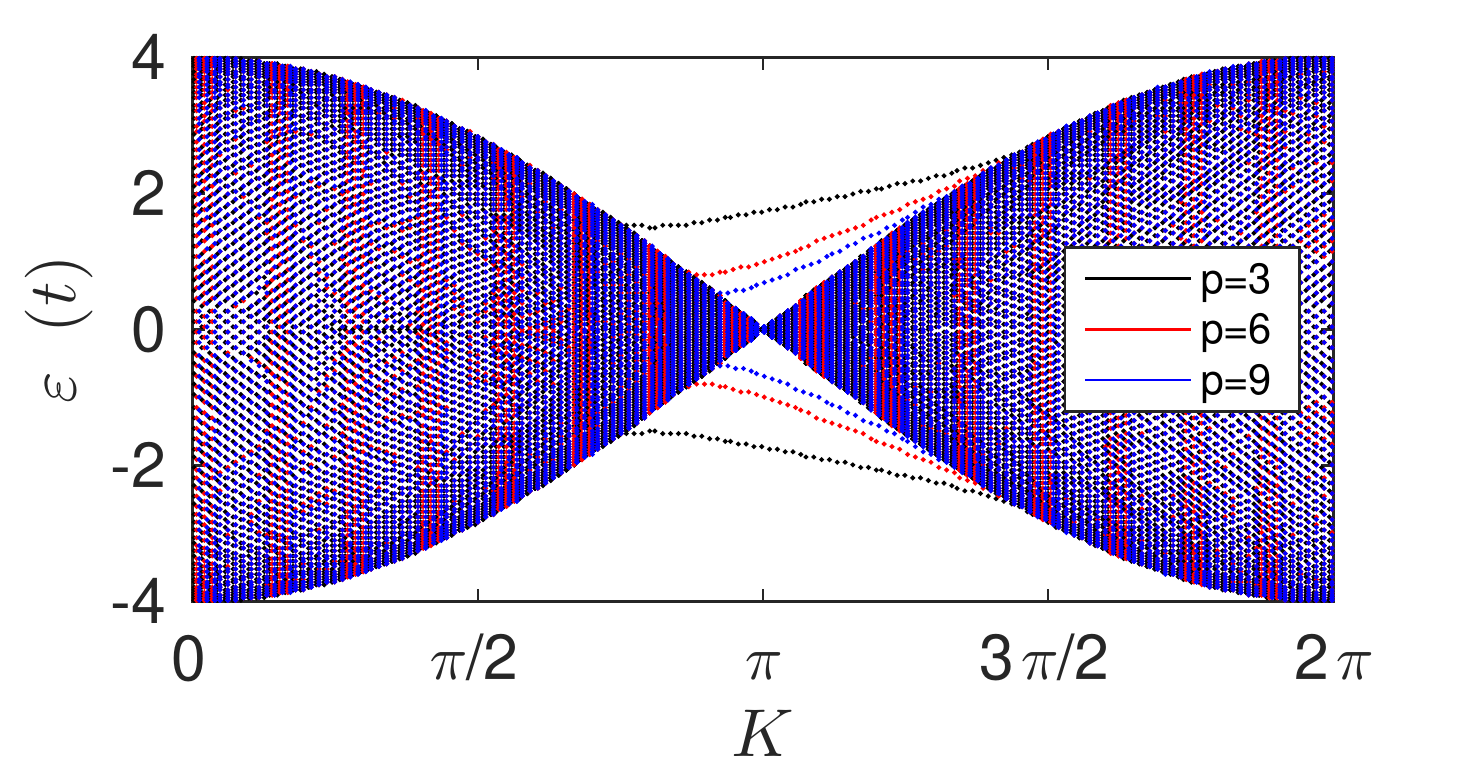}
  \caption{Two-body spectrum (data points) for a tight-binding model of FPs with $L=151$ sites, PBC, and $p=3$ (upper and lower panel),
    $p=6$, $p=9$ (lower panel), plotted versus the center-of-mass momentum $K$.
    In the upper panel, data points evidenced in magenta correspond to bound states, while all the others are associated to extended states.
    The arrows (resp., the square) denote the various values of $K$ associated to bound states
    (resp., to the scattering state) whose weight distribution is displayed in Fig.~\ref{Fig:Weight_k}.}
  \label{Fig:TwoBody}
\end{figure}
%%%%%%%%%%%%%%%%%%%%%%%%%%%%%%%%%%%%%%%%%%%%%%%%%%%%%%%%%%%%%%%%%%%%%%%%%%%%%%%%

\begin{figure}[!t]
  \includegraphics[width=\columnwidth]{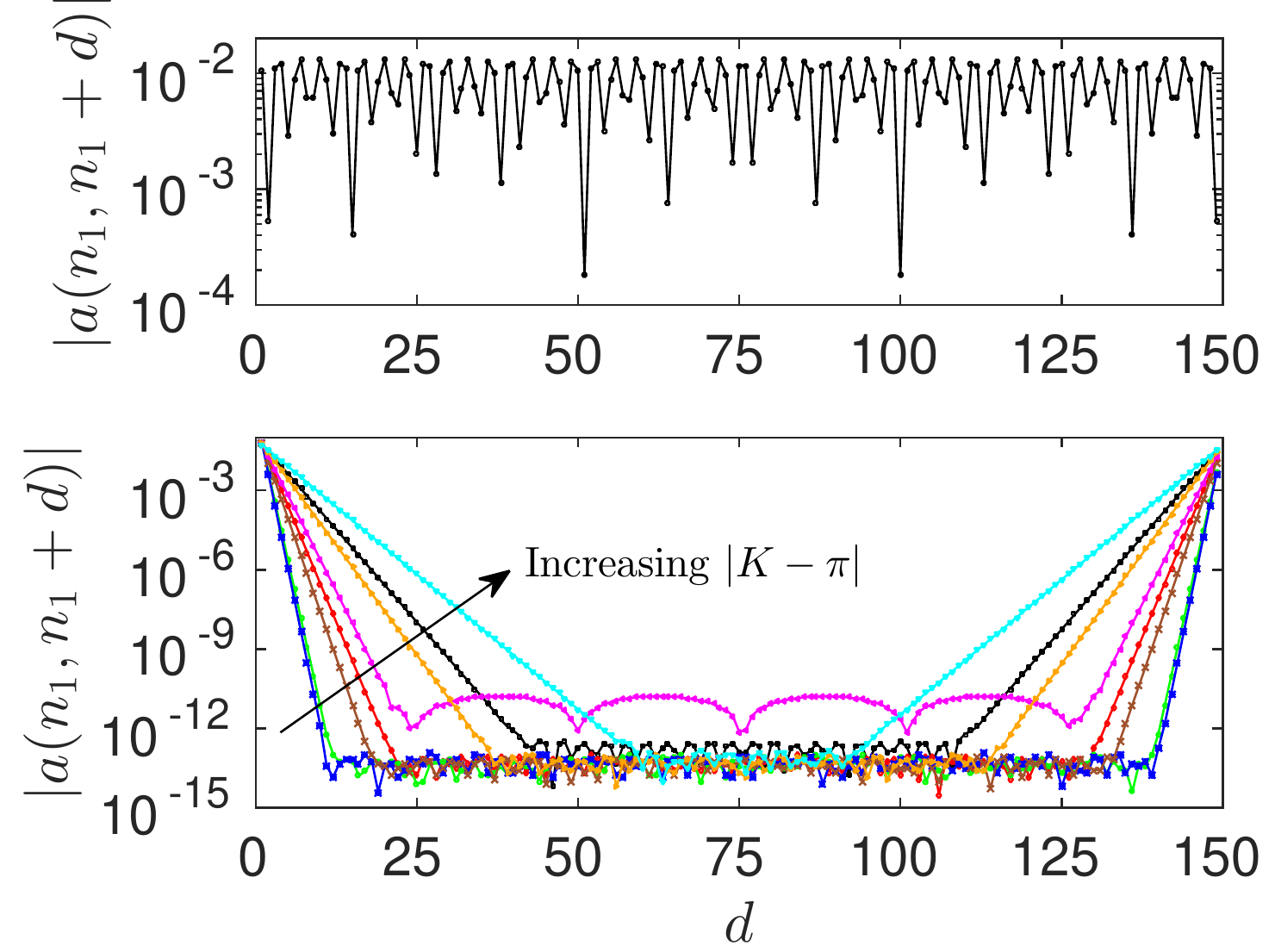}
  \caption{Absolute value of the weight distribution $|a(n_1, n_2)|$ as a function of the distance
    $d= n_2-n_1$, for various eigenstates of the two-body tight-binding FPs Hamiltonian ($p=3$).
    Upper panel: weight distribution for the scattering state evidenced by a square
    in Fig.~\ref{Fig:TwoBody}. Lower panel: weight distributions for the bound states
    denoted by arrows in Fig.~\ref{Fig:TwoBody}; the closer the states are to the two lobes,
    the largest is the width of the distribution.}
  \label{Fig:Weight_k}
\end{figure}

In Fig.~\ref{Fig:TwoBody} we show the two-body spectrum for a system of $L=151$ sites, $p=3$ (upper and lower panel),
$p=6$, and $p=9$ (lower panel), obtained by means of an exact diagonalization (ED) of Eq.~\eqref{Eq:Ham} with PBC.
Most of the spectrum is composed by two-parafermion scattering states, for which an analytical solution of the form~\cite{Karbach_1997}
\begin{equation}
  a(n_1, n_2) = \left\{
  \begin{matrix}
    A \left( e^{i (k n_1+q n_2)} + e^{i \theta } e^{i (q n_1+k n_2)} \right)  & n_1 < n_2, \vspace{0.5mm} \\
    B e^{i (k+q) n_1} & n_1 = n_2,
  \end{matrix} \right.
  \label{Eq:2bodyAnsatz2}
\end{equation}
with $q,k \in \mathbb R$, can be obtained.
In Fig.~\ref{Fig:Weight_k}, upper panel, we show $|a(n_1, n_2)|$ for the scattering state highlighted by a square
in Fig.~\ref{Fig:TwoBody} (upper panel, for $p=3$), which is clearly delocalized over the full length of the system
[note that for the states in~\eqref{Eq:2bodyAnsatz2}, the center-of-mass momentum is given by $K = k+q$].

To further gain analytical insight in the physics of the scattering states, we observe that,
because of the anyonic statistics, once we impose PBC, we obtain $a (n_1, n_2) = \omega^* a(n_2, n_1+L)$.
This leads to the set of equations:
\begin{subequations}
  \label{Eq:kqQuant}
  \begin{align}
    k =& \frac{2 \pi }{L} \left( \lambda_k - \frac{1}{p}\right) - \frac{\theta}{L},
    \quad \rev{\lambda_k \in \{0,1, \ldots, L-1 \}},\\
    q =& \frac{2 \pi }{L} \left( \lambda_q - \frac{1}{p}\right) + \frac{\theta}{L},
    \quad \rev{\lambda_q \in\{0,1, \ldots, L-1 \}}. 
  \end{align}
\end{subequations}
As already pointed out in 1D anyonic models that are solvable through Bethe ansatz,
the momenta are shifted by a quantity which is proportional to the statistical parameter, namely $\kappa \pi /L$~\cite{Calabrese_2007}.
The value of the phase $\theta$ can be determined numerically by solving the equations obtained
by projecting the eigenvalue equation $\hat H \ket{\Psi} = E \ket {\Psi}$
over a state $\bra{n_1, n_2}$ (more details are given in App.~\ref{App:2body}).
We have cross-checked that all the obtained eigenenergies are reproduced by ED calculation.

Comparing with typical two-body spectra of noninteracting 1D quantum systems,
it is immediate to recognize that there are states whose energy behaves differently from
the two-parafermion scattering states (see the two mustaches in Fig.~\ref{Fig:TwoBody}, upper panel,
highlighted in magenta). They correspond to bound states because the ratio of the amplitudes for closer
and separated particles is large: $a(n_1, n_1) / a(n_1, n_1+L/2)\gg 1$~\cite{Karbach_1997}.
This is only possible if $k$ and $q$ in wavefunction~\eqref{Eq:2bodyAnsatz} have an imaginary part.
A closer inspection at the full weight distribution $a(n_1, n_2)$ for bound states indeed shows that
it decays exponentially fast with the distance $d = n_2-n_1$, and is peaked at $n_2=n_1$ (see Fig.~\ref{Fig:Weight_k}, lower panel).
The width depends on the imaginary parts of $k$, $q$, $\theta$,
and reaches its minimum for $k+q=\pi$.
For states departing from this condition, but still in the mustache,
the binding of the two particles loosens, although it remains exponential.
Conversely, for states belonging to the two lobes of Fig.~\ref{Fig:TwoBody},
the weight distribution is delocalized over all the chain, thus signaling scattering states
(see Fig.~\ref{Fig:Weight_k}, upper panel).
Finally we mention that, as expected, by increasing $p$, the mustaches
of bound states become less visible and merge into the continuum of scattering states (see Fig.~\ref{Fig:TwoBody}, lower panel).
Indeed the width of distribution $|a(n_1, n_1+d)|$ for bound states
progressively increases toward an extended configuration [not shown].

%%%%%%%%%%%%%%%%%%%%%%%%%%%%%%%%%%%%%%%%%%%%%%%%%%%%%%%%%%%%%%%%%%%%%%%%%%%%%%%%
%%%%%%%%%%%%%%%%%%%%%%%%%%%%%%%%%%%%%%%%%%%%%%%%%%%%%%%%%%%%%%%%%%%%%%%%%%%%%%%%

\section{Level spacing statistics and Integrability}\label{Sec:LSS}

To corroborate the statement mentioned in Sec.~\ref{Sec:Relation} that the Hamiltonian~\eqref{Eq:Ham}, for $p>2$,
is not integrable and does not enjoy Bethe-ansatz solvability, we have studied its LSS.
The statistics of the energy eigenstates of $\hat H$, being a key feature of the spectrum
of a generic quantum system, represents the standard tool to investigate its possible integrability~\cite{Haake}.
As a matter of fact, the key feature of integrable systems is a tendency of levels to cluster
and eventually cross when a given Hamiltonian parameter is varied,
due to the presence of a number of integrals of motion. Conversely, in non-integrable systems,
the absence of non-trivial conserved laws correlates the levels in such a way to avoid crossings.

To quantitatively characterize these tendencies, it is useful to analyze the
probability distribution $P(s)$ that the energy difference
between two adjacent levels $s_n = E_{n+1} - E_n$ (normalized to
the average level spacing) lies in a given interval $[s,s + ds]$.
For integrable systems one typically obtains a Poissonian (P) statistics,
\begin{equation}
  P_{\rm P}(s)=e^{-s},
\end{equation}
as usual for uncorrelated levels coming from different symmetry sectors.
For non-integrable systems the spectrum is conjectured to follow
the rules of random matrix theory, leading to a Wigner-Dyson (WD) surmise,
\begin{equation}
  P_{\rm WD}(s) \sim A s^\beta e^{-B s^2},
\end{equation}
where level repulsion manifests in the fact that
\begin{equation}
  \lim_{s\to 0} P_{\rm WD}(s) \sim s^\beta , \qquad \beta>0.
  \label{eq:LevRep}
\end{equation}
More in details, depending on the symmetries of the corresponding Hamiltonian,
the WD distribution presents a specific shape; for example, for systems 
preserving one anti-unitary symmetry (e.g., invariance under time-reversal),
the LSS is given by a Gaussian orthogonal ensemble (GOE), where
$P_{\rm GOE} = \tfrac{\pi s}{2}e^{-\pi s^2/4}$, with $\beta = 1$.
Under more general conditions, the LSS of complex Hamiltonians is generally captured by
a Gaussian unitary ensemble (GUE), such that
$P_{\rm GUE} = \tfrac{32 s^2}{\pi^2} e^{-4 s^2/\pi}$, with $\beta = 2$.

In our case, by means of ED, we have checked that the spectrum
of the FP tight-binding Hamiltonian systematically develops level repulsion.
In order to avoid any effect of level crossings due to trivial symmetries,
we have numerically studied the full spectrum of Eq.~\eqref{Eq:Ham} for a fixed number of particles,
and with open boundary conditions (OBC).
In computing the LSS, we have also dropped the lower and upper third of the energy levels,
since generic non-integrable systems typically exhibit level repulsion only in the central band of the spectrum.

%%%%%%%%%%%%%%%%%%%%%%%%%%%%%%%%%%%%%%%%%%%%%%%%%%%%%%%%%%%%%%%%%%%%%%%%%%%%%%%%
\begin{figure}[!t]
  \includegraphics[width=0.9\columnwidth]{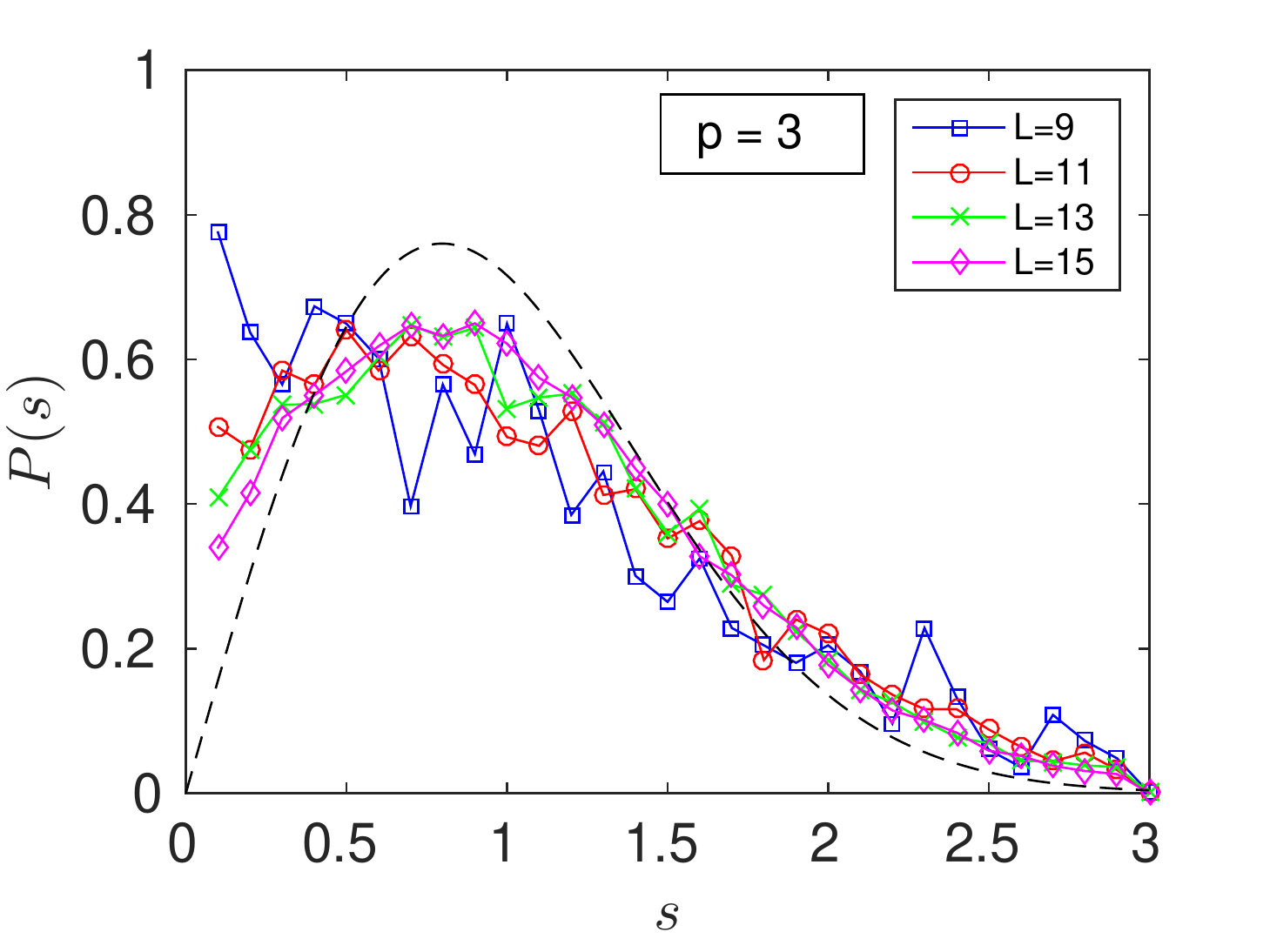}
  \includegraphics[width=0.9\columnwidth]{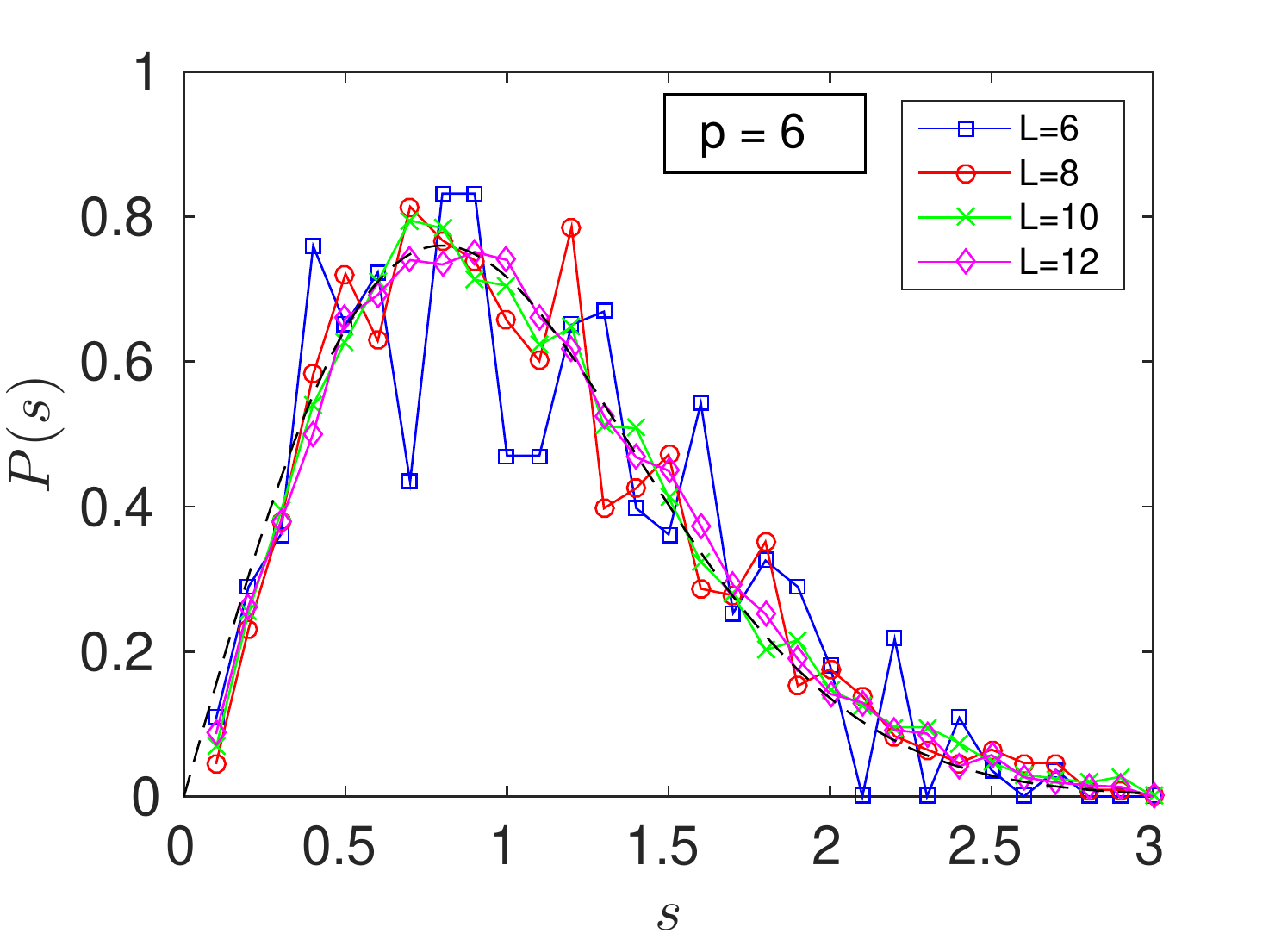}
  \caption{LSS for the FP tight-binding model of Eq.~\eqref{Eq:Ham},
    with $p=3$ (upper panel) and $p=6$ (lower panel), for a fixed number of $N=7$ particles
    and different chain lengths, as specified in the legend.
    The dashed curve corresponds to a GOE statistics. OBC have been enforced.
    For the largest available system sizes, the statistics is performed over the central
    $\sim 2 \times 10^{4}$ energy levels.}
  \label{fig:LSS}
\end{figure}
%%%%%%%%%%%%%%%%%%%%%%%%%%%%%%%%%%%%%%%%%%%%%%%%%%%%%%%%%%%%%%%%%%%%%%%%%%%%%%%%

Results for chains of various size are displayed in Fig.~\ref{fig:LSS},
for a fixed number of $N=7$ FPs corresponding to $p=3$ (upper panel) and $p=6$ (lower panel).
While at small lengths the LSS displays a rather irregular pattern, when increasing $L$
we observe a clear tendency to develop a peak at intermediate values of $s$, thus evidencing
the behavior in Eq.~\eqref{eq:LevRep}, typical for non-integrable models.
More specifically, for the sizes we were able to reach, at $p=6$ the distribution $P(s)$
exhibits a fast convergence to a GOE surmise (lower panel); at $p=3$ the situation is less clear
and larger sizes would be required (upper panel).
We have checked that the above scenario is not affected by the choice of $N$ [data not shown].
It is also worth mentioning that the asymptotic WD distribution to which the LSS of the FP spectrum converges, 
is expected to depend on the specific symmetries of $\hat H$, as detailed in App.~\ref{App:LSS_Fermi}.

Finally we wish to stress that the $p=2$ case, in which FPs turn out to be canonical fermions,
is different in this respect, since it can be trivially integrated in momentum space.
This reflects into a Poissonian LSS (see App.~\ref{App:LSS_Fermi}).

\section{Many-body physics}\label{Sec:Z3}

We now move to the study of many-body properties of the model, explicitly focusing on the $p=3$ and $p=6$ cases.
For each of them, we consider values of the density which satisfy $0 < N/L \leq 1$.
Notice that, for $p=3$, these results span all the possibilities, since densities larger than $1$
are unitarily equivalent to smaller ones (see Sec.~\ref{Sec:Ham:Sym}).
In all situations, we have employed a DMRG-based numerical approach~\cite{Schollwoeck_2005}.
Specifically, we have simulated systems with up to $L=288$ sites, OBC, and several particle numbers ranging between $N=24$ and $N=288$.
The number of kept states is $m \leq 250$, such that the truncation error is always smaller than $10^{-8}$.
The simulations are performed by applying the Fradkin-Kadanoff transformation~\cite{Fradkin_1980} to the model,
so that it is defined in terms of more conventional commuting operators (see App.~\ref{App:FKtrasf} for details).

\subsection{Low-energy properties}
\label{subsec:Low-ener}

\begin{figure}[!t]
 \includegraphics[width=0.8\columnwidth]{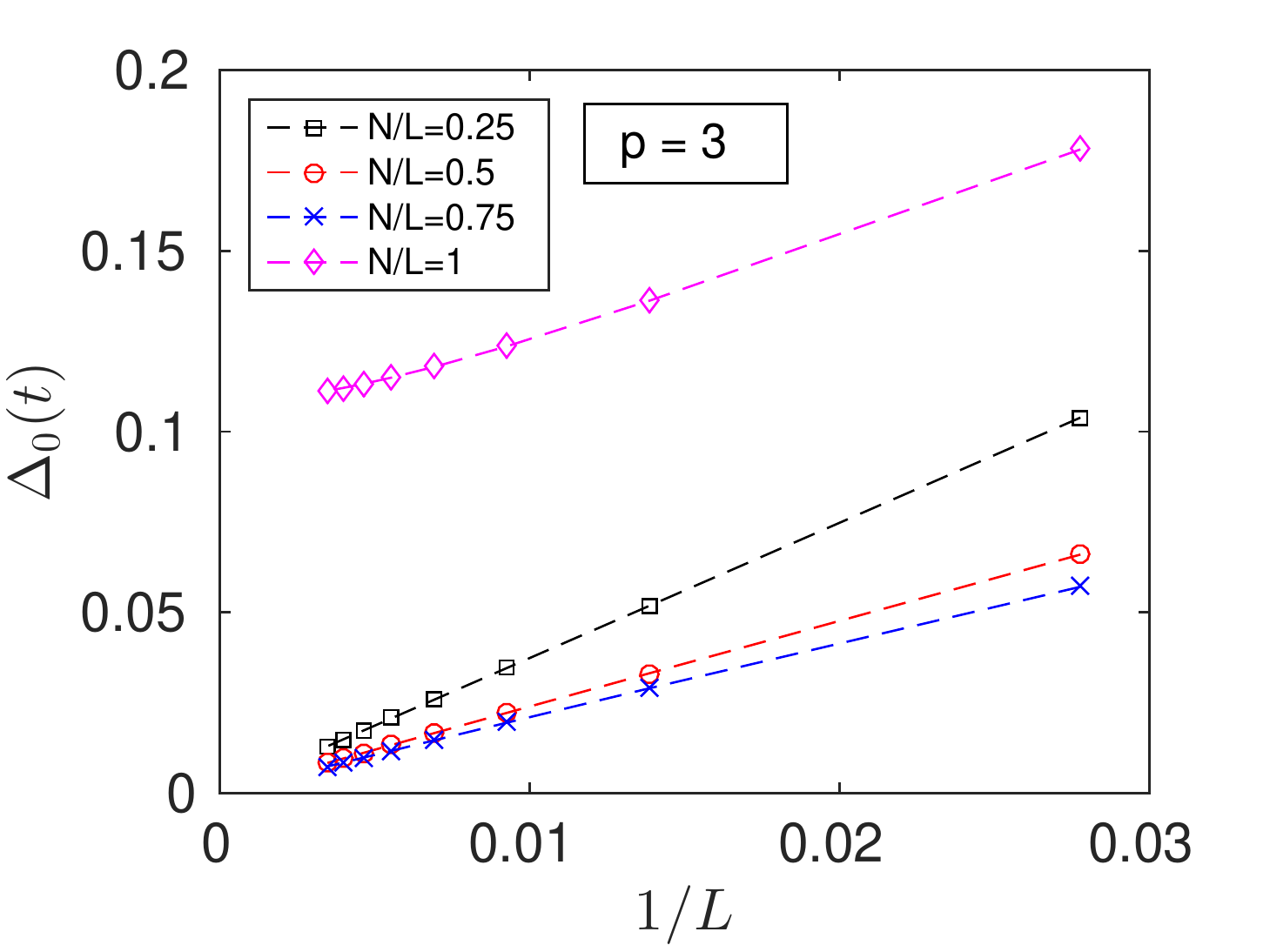}
 \includegraphics[width=0.8\columnwidth]{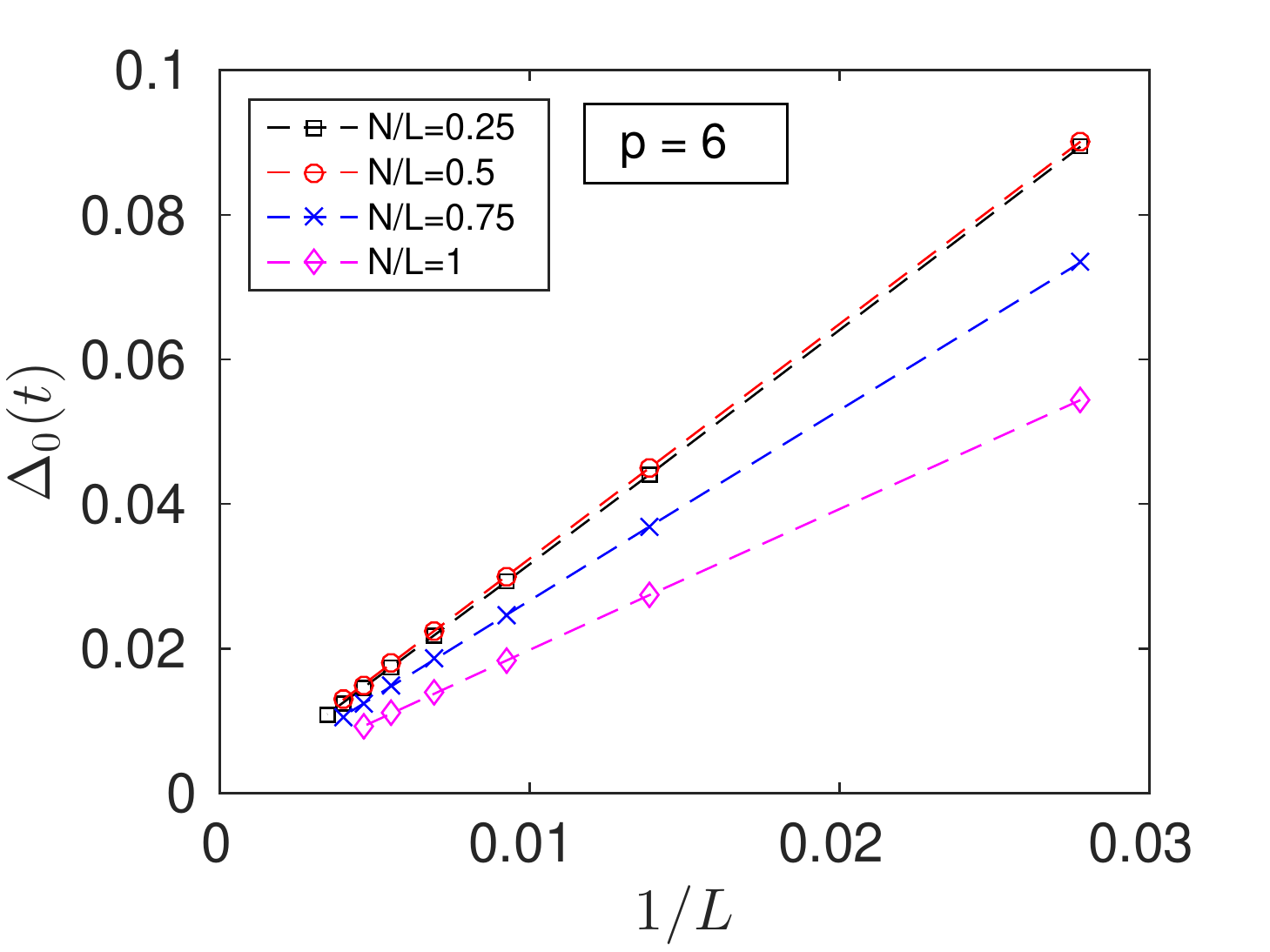}
 \caption{Energy gap between the first excited state and the ground state for a fixed number of particles
   in the case $p=3$ (upper panel) and $p=6$ (lower panel) plotted versus $L^{-1}$.
   The maximal system length considered is $L=288$.}
 \label{Fig:Z3:Z6:Gap}
\end{figure}

Let us start our many-body analysis by focusing on the lower part of the spectrum of Hamiltonian~\eqref{Eq:Ham}.
We first compute the neutral gap $\Delta_0$ of the system, namely the energy difference between the first excited state
and the ground state for a fixed number $N$ of particles. Results are shown in Fig.~\ref{Fig:Z3:Z6:Gap}
for several values of $N/L$, in the cases of $p=3$ (upper panel) and of $p=6$ (lower panel). 
For $N/L=1$ and $p=3$, we observe the opening of an energy gap, while in all other cases the gap closes as $L^{-1}$.
The latter behavior is the unambiguous hallmark of an approximate low-energy conformal invariance.
We thus expect the system to be generally described, at low energies, by a conformal field theory (CFT).

%%%%%%%%%%%%%%%%%%%%%%%%%%%%%%%%%%%%%%%%%%%%%%%%%%%%%%%%%%%%%%%%%%%%%%%%%%%%%%%%
%%%%%%%%%%%%%%%%%%%%%%%%%%%%%%%%%%%%%%%%%%%%%%%%%%%%%%%%%%%%%%%%%%%%%%%%%%%%%%%%

%%%%%%%%%%%%%%%%%%%%%%%%%%%%%%%%%%%%%%%%%%%%%%%%%%%%%%%%%%%%%%%%%%%%%%%%%%%%%%%%
\begin{figure}[!t]
 \includegraphics[width=0.8\columnwidth]{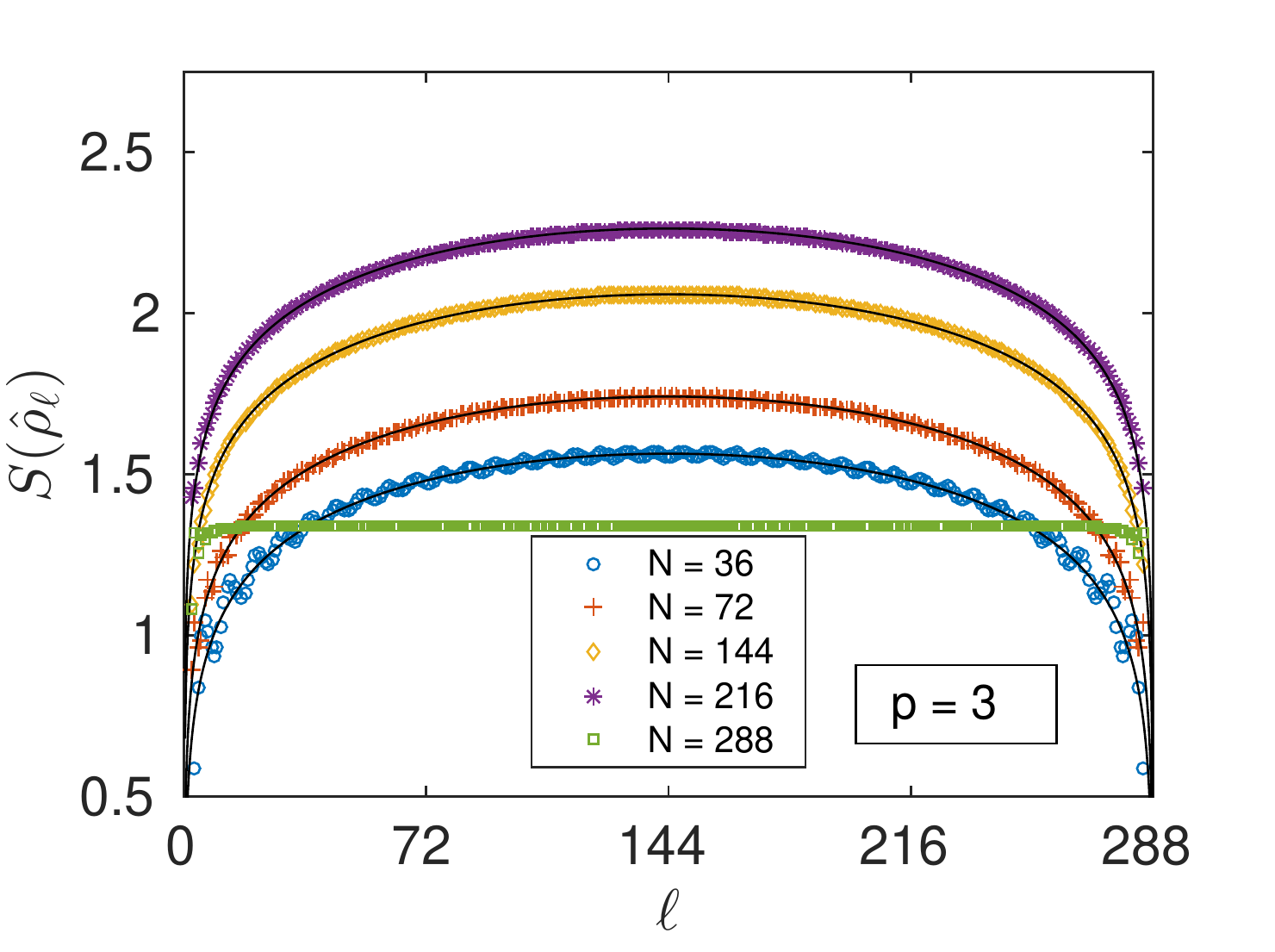}
 \includegraphics[width=0.8\columnwidth]{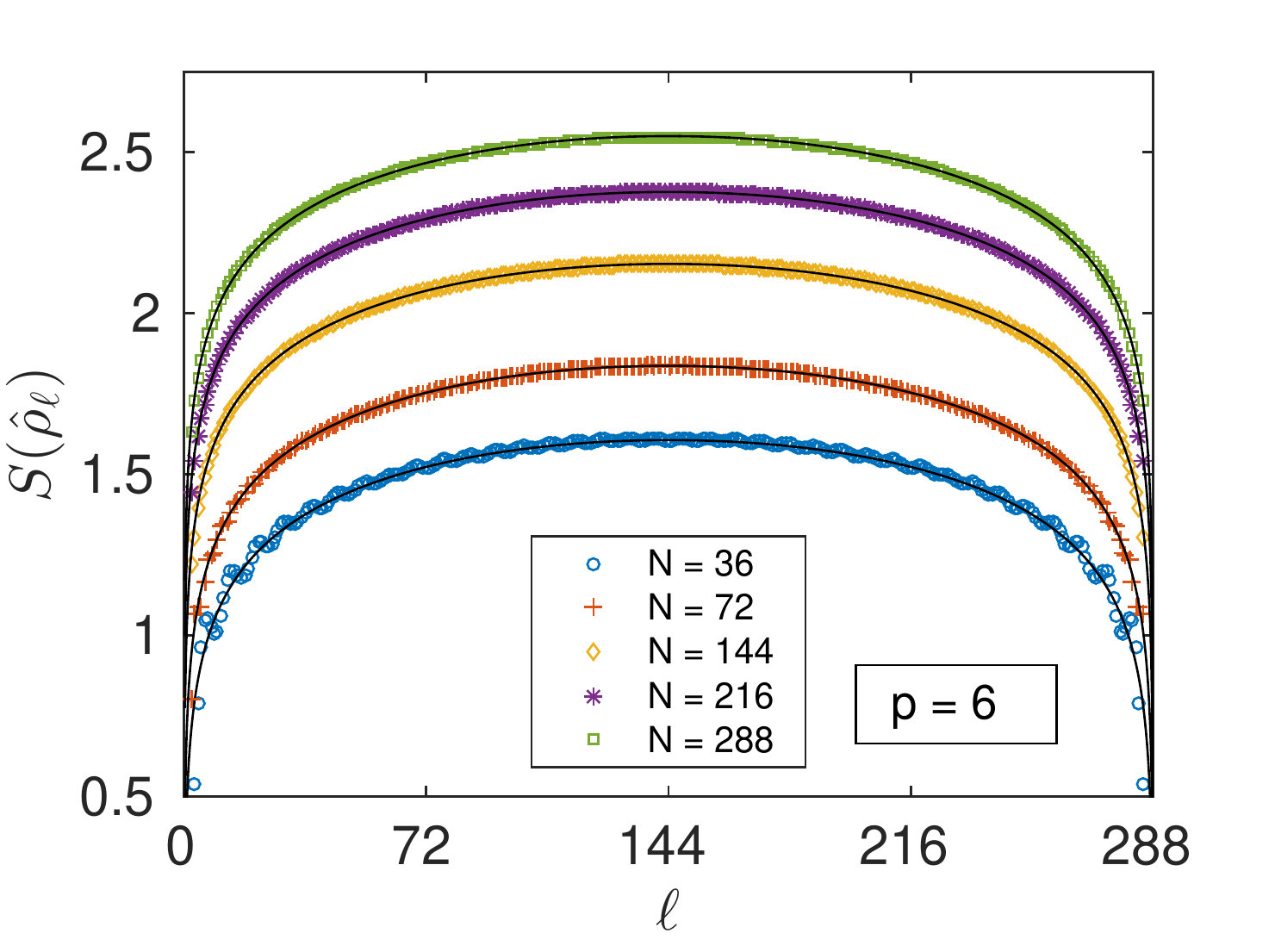}
 \caption{Von Neumann entropy $S(\hat \rho_{\ell})$ for several particle numbers and $L=288$,
   in the cases $p=3$ (upper panel) and $p=6$ (lower panel).
   Thin black lines are the fitting curves of the numerical data (symbols),
   as obtained using the formula in Eq.~\eqref{Eq:Calabrese:Cardy} with $c=1$ and $a$ left as a fit parameter.}
 \label{Fig:Z3:Z6:VN}
\end{figure}
%%%%%%%%%%%%%%%%%%%%%%%%%%%%%%%%%%%%%%%%%%%%%%%%%%%%%%%%%%%%%%%%%%%%%%%%%%%%%%%%

To further assess the low-energy properties of the system, we have also calculated the 
bipartite entanglement entropy of the ground state $\ket{\Psi_{\rm GS}}$. This quantifies the amount
of genuine quantum correlations that establish among two parts of a given bipartition of the system,
that is, between the first $\ell$ and the last $L-\ell$ sites.
After taking the reduced density matrix of the first part
\begin{equation}
  \hat \rho_\ell = \text{Tr}_{L-\ell} \, \big[ \ket{\Psi_{\rm GS}} \bra{\Psi_{\rm GS}} \big] ,
\end{equation}
the entanglement of the bipartition is defined through the so-called von Neumann entropy
\begin{equation}
  S(\hat \rho_\ell) = - \text{Tr} \, \big[ \hat \rho_\ell \, \log (\hat \rho_\ell) \big].
\end{equation}
For the ground state of a 1D CFT, this can be shown to behave as:
\begin{equation}
 S(\hat \rho_\ell) = a + \frac{c}{6} \log \left[ 
 \frac{2 L}{\pi} \sin \left( \frac{\pi \ell}{L} \right)  \right],
 \label{Eq:Calabrese:Cardy}
\end{equation}
where $c$ denotes the central charge of the theory~\cite{Calabrese_2009}.

The outcomes of our DMRG computations for the entanglement entropy are reported in Fig.~\ref{Fig:Z3:Z6:VN}.
We have fitted the numerical data (symbols) with the formula in Eq.~\eqref{Eq:Calabrese:Cardy}, setting $c=1$
and leaving $a$ as the only fit parameter. As is clearly visible from the figure, the agreement is extremely good
and certifies that the low-energy theory of Hamiltonian~\eqref{Eq:Ham} is a CFT with $c=1$.
As such, the model is amenable to a low-energy description in terms of a Luttinger liquid (LL) (see Sec.~\ref{Sec:Fermi}).
We point out that, as expected, in the gapped case ($p=3$ and $N/L=1$) the entanglement entropy does not follow
the scaling in Eq.~\eqref{Eq:Calabrese:Cardy}, while rather it satisfies an area law, namely it saturates
to a finite value without diverging with $\ell$ (green data set in upper panel of Fig.~\ref{Fig:Z3:Z6:VN}).

The appearance of a gapped phase at commensurate density in the tight-binding model for $p=3$
is a peculiarity of the anyonic statistics.
Whereas in bosonic models the system is always gapless, for spinless fermions it would be a trivial band insulator,
since in that case the system is completely filled. Conversely, in the case of spin-1/2 fermions (e.g. electrons),
at $N/L=1$ the system remains gapless if quartic terms are disregarded. As such, for $p=3$ and $N/L=1$, the system
is in an anyonic Mott-like phase (the concept of band insulator is not easily generalizable to anyons)
related to non-linearities of anyonic definition. 
It is interesting to note that, contrary to what happens here, in the lattice anyon-Hubbard model
there is no gapped phase at commensurate fillings, in the absence of quartic terms.

\subsection{Anyonic correlation functions in the gapless cases}
\label{subsec:gapless}

We now move to the study of some relevant observables for our anyonic gas.
The density profile does not display any exotic property, and it resembles in several respects
that of a gas of repelling particles confined in 1D. We observed the presence of Friedel-like oscillations
with a space period equal to the inverse density $L/N$ [not shown]. 

\begin{figure}[!t]
 \includegraphics[width=0.8\columnwidth]{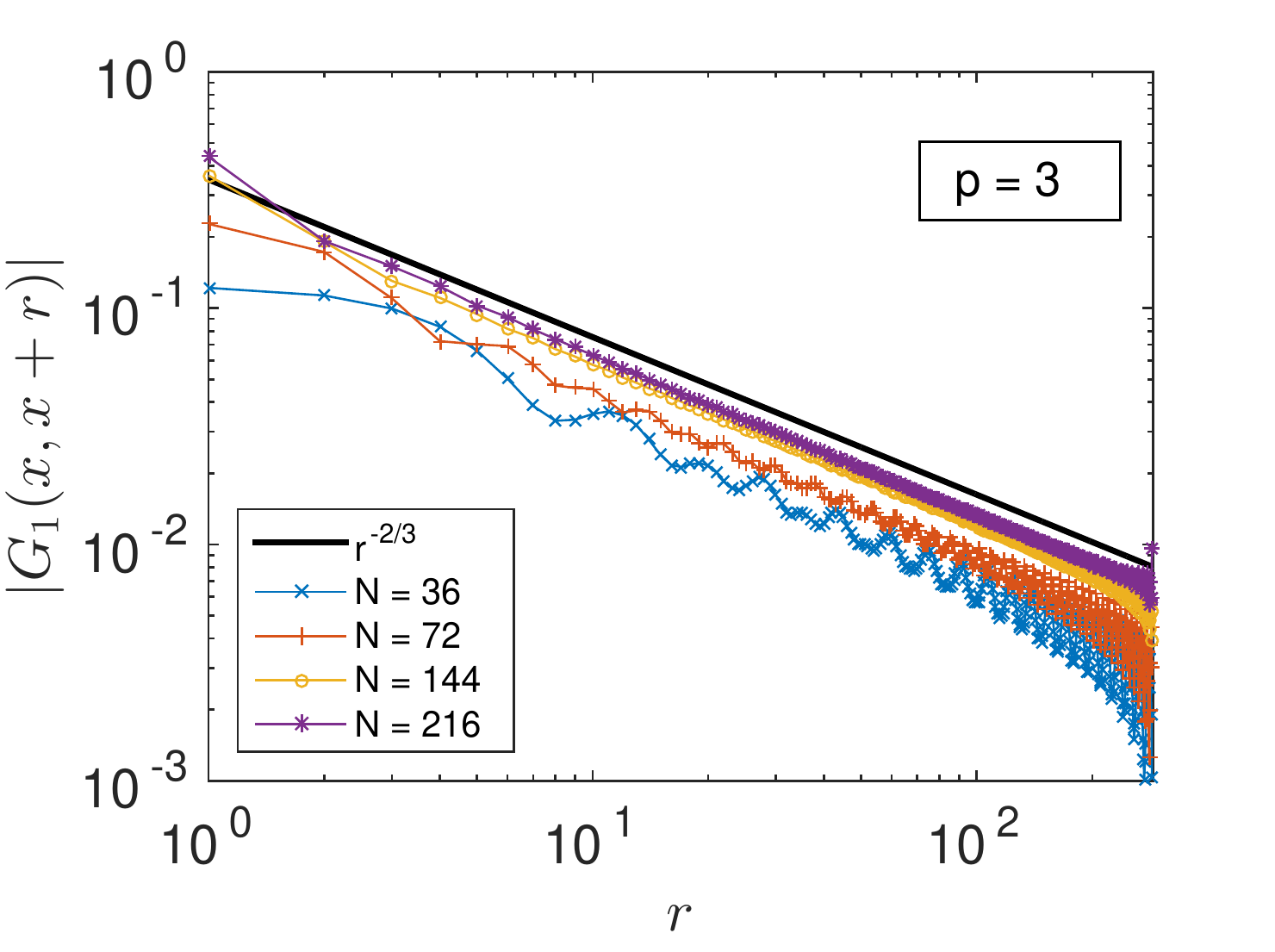}
 \includegraphics[width=0.8\columnwidth]{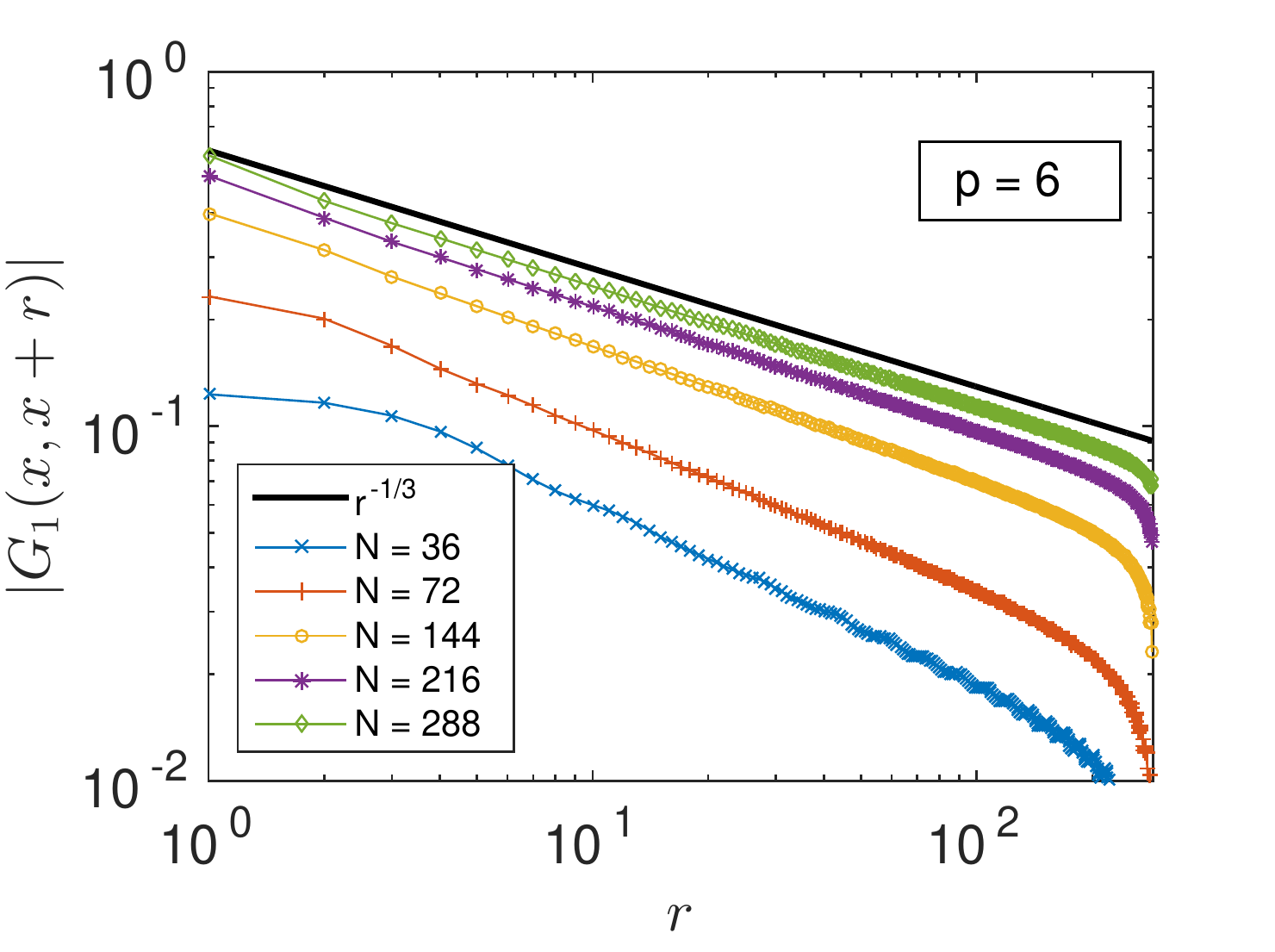}
 \caption{Absolute value of the one-body density matrix $G_1(x,x+r)$ as a function of $r$
   in the cases $p=3$ (upper panel) and $p=6$ (lower panel), for several particle numbers
   and fixed chain length $L=288$.
   Thick black lines in the two panels indicate the predictions of Refs.~\cite{Calabrese_2007, Bellazzini_2009}.
   Data points (symbols) have been fitted with a power-law function $|G_1| \sim r^{-\alpha_1}$, in the range $r \in [10, L/2]$.
   For $p=3$, the resulting best-fit value of the exponent is:
   $\alpha_1 = 0.74 \pm 0.03$ ($N=36$), $\alpha_1 = 0.70 \pm 0.01$ ($N = 72$), $\alpha_1 = 0.675 \pm 0.001$ ($N = 216$).
   For $p=6$, we get: $\alpha_1 = 0.460 \pm 0.002$ ($N=72$), $\alpha_1 = 0.385 \pm 0.001$ ($N = 144$),
   $\alpha_1 = 0.353 \pm 0.001$ ($N = 216$), $\alpha_1 = 0.341 \pm 0.001$ ($N = 288$).}
  \label{Fig:Z3:Z6:G1}
\end{figure}

As we shall see below, the two-point correlation functions will reveal more insightful quantities.
Let us first analyze the one-body density matrix
\begin{equation}
  G_1(j,l) = \langle \Psi_{\rm GS} | \hat F_j^\dagger \, \hat F_l | \Psi_{\rm GS} \rangle.
  \label{eq:G1}
\end{equation}
Since we are using OBC, in order to minimize boundary effects, we measure correlations between two points
that are symmetrically chosen with respect to the center of the chain.
Figure~\ref{Fig:Z3:Z6:G1} shows the absolute value $|G_1(x,x+r)|$ as a function of the distance $r$,
for $p=3$ (upper panel) and $p=6$ (lower panel), and for several values of $N/L$ such that the ground-state energy
gap vanishes in the thermodynamic limit.
A clear power-law decay $r^{-\alpha_1}$ emerges, consistently with the fact that the phase is gapless.
We observe that, whereas in the $p=3$ situation the fitted exponent $\alpha_1$ is approximately
the same in the wide range of densities between $N/L = 1/8$ and $3/4$, more differences appear
in the case $p=6$ (see the caption of Fig.~\ref{Fig:Z3:Z6:G1} for the extrapolated values of $\alpha_1$).

%%%%%%%%%%%%%%%%%%%%%%%%%%%%%%%%%%%%%%%%%%%%%%%%%%%%%%%%%%%%%%%%%%%%%%%%%%%%%%%%
%%%%%%%%%%%%%%%%%%%%%%%%%%%%%%%%%%%%%%%%%%%%%%%%%%%%%%%%%%%%%%%%%%%%%%%%%%%%%%%%

It is instructive to compare our numerical data with previously-developed analytical results
for anyonic gases. In particular, we now try to match them with those of Refs.~\cite{Calabrese_2007, Bellazzini_2009},
providing a description of correlation functions for a 1D anyonic gas, based on an effective low-energy LL description.
Let us however stress that it is not \textit{a priori} clear that such description is applicable to our model, since the former
is developed by deforming bosonic field operators into anyonic ones and the second is simply introduced as a continuum anyonic model.
We now assess whether the predictions of Refs.~\cite{Calabrese_2007, Bellazzini_2009} in the case
of non-interacting anyons describe our model.
The correlation function of Eq.~\eqref{eq:G1} is predicted to scale as $|G_1(x,x+r)| \sim r^{-(\kappa^2 K+ 1/K)/2}$,
where $K$ is the Luttinger parameter. Since in the anyonic Luttinger model $K = \kappa^{-1}$, it follows that
in our case $\alpha_1 = \kappa = 2/p$. This prediction is indicated in Fig.~\ref{Fig:Z3:Z6:G1} with a thick black line.
The comparison with the fitted values of $\alpha_1$ improves when increasing the density $N/L$.
We thus conclude that our model, in the gapless region, is well approximated by the universal LL description
proposed in Refs.~\cite{Calabrese_2007, Bellazzini_2009} for the non-interacting anyonic gas.

In passing we mention that, for the special case at unit filling and $p=3$, where a gapped phase develops
(see Sec.~\ref{subsec:Low-ener}), correlations functions develop important qualitative differences.
Specifically, as one should expect, the one-body density matrix decays exponentially as $e^{- r/\xi}$,
$\xi$ being the proper (finite) correlation length [not shown].

%%%%%%%%%%%%%%%%%%%%%%%%%%%%%%%%%%%%%%%%%%%%%%%%%%%%%%%%%%%%%%%%%%%%%%%%%%%%%%%%
\begin{figure}[t]
 \includegraphics[width=0.8\columnwidth]{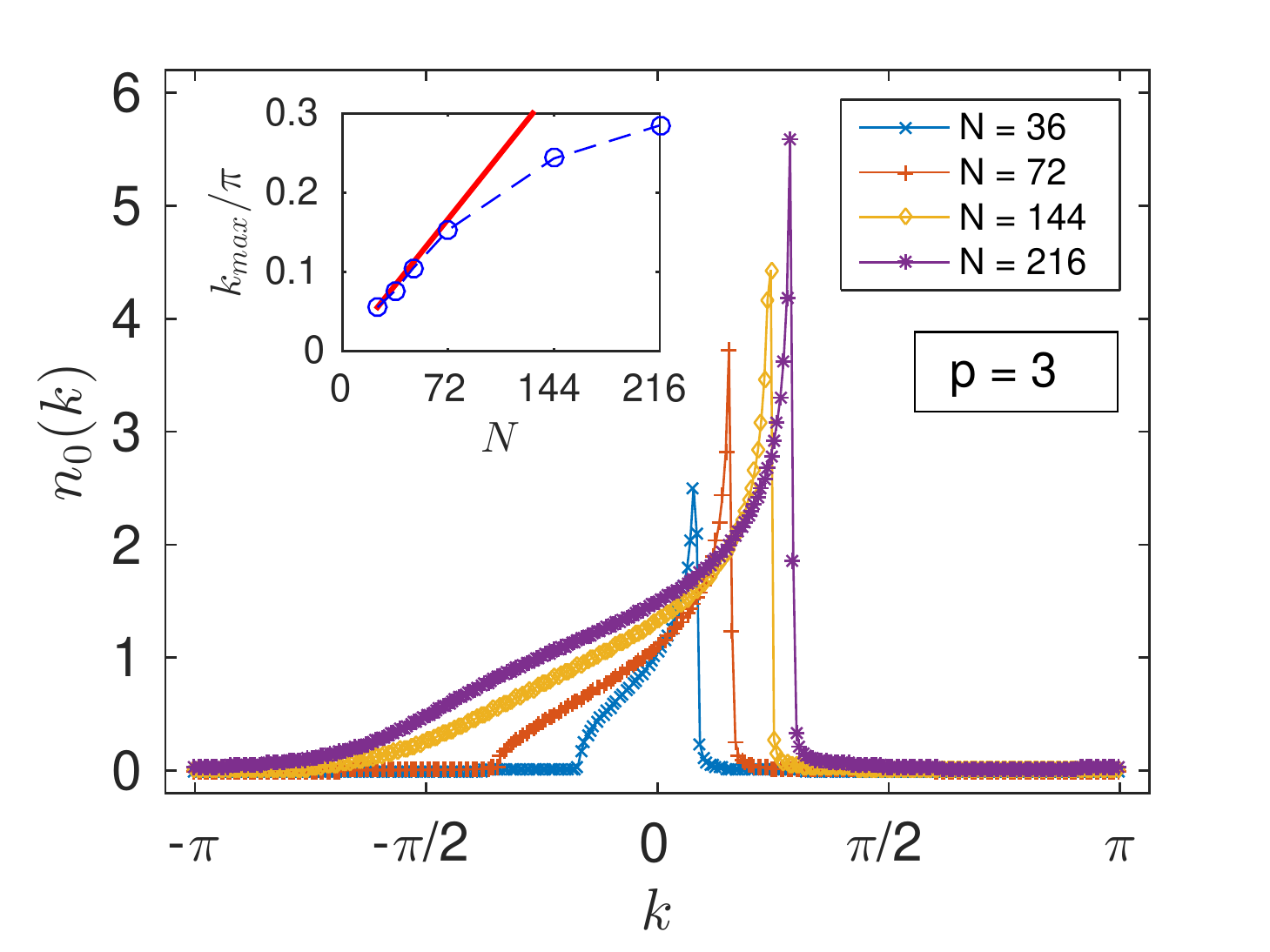}
 \includegraphics[width=0.8\columnwidth]{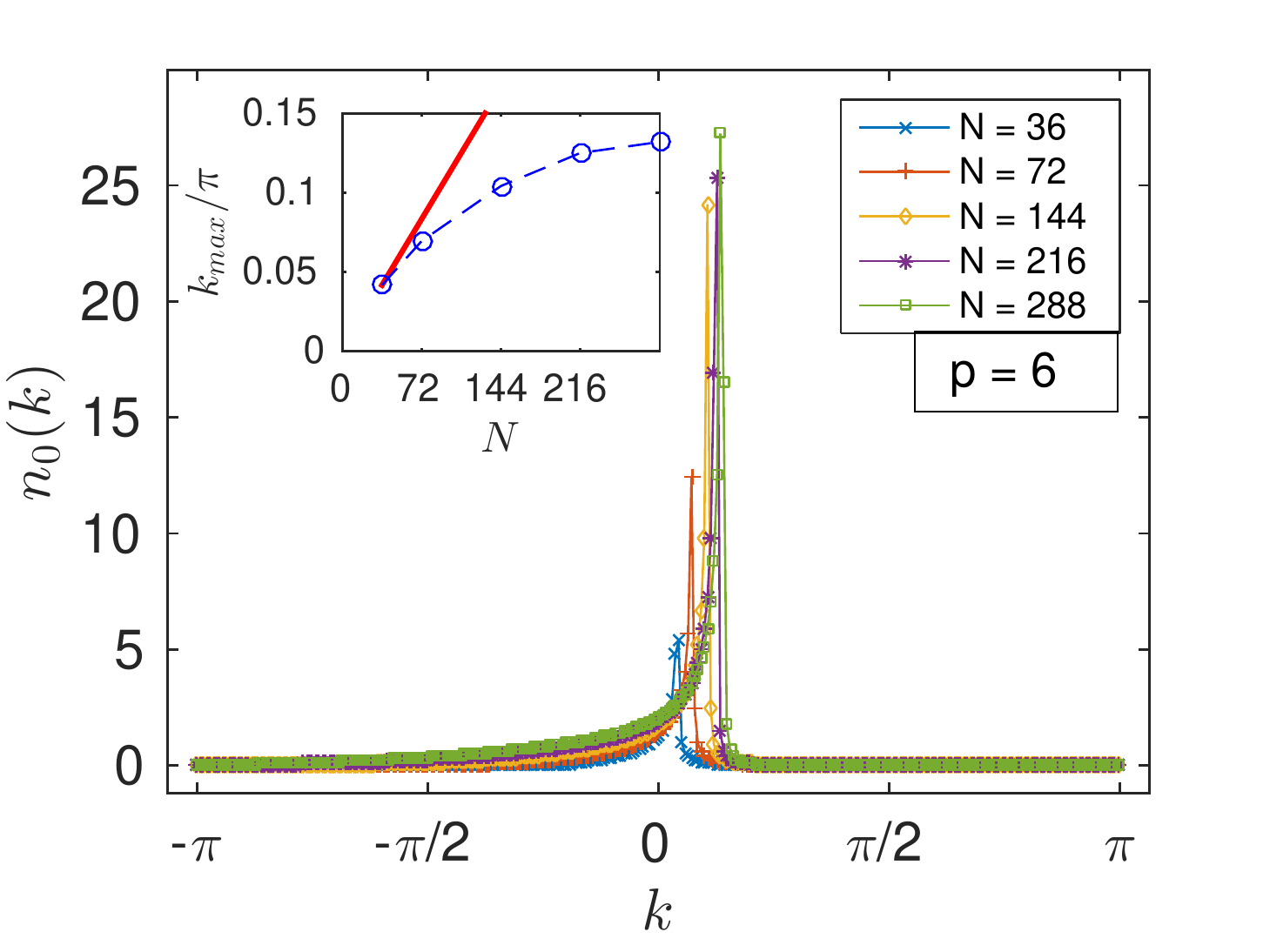}
 \caption{Momentum distribution function $n_0(k)$ for $p=3$ (upper panel) and $p=6$ (lower panel),
   for several particle numbers and $L=288$.
   Inset: plot of the peak position as a function of $N$.
   The red line denotes the position $k_{\rm max}$ discussed in the text.}
 \label{Fig:Z3:Z6:nk}
\end{figure}
%%%%%%%%%%%%%%%%%%%%%%%%%%%%%%%%%%%%%%%%%%%%%%%%%%%%%%%%%%%%%%%%%%%%%%%%%%%%%%%%

Differently from what happens in the bosonic and fermionic version of Hamiltonian~\eqref{Eq:Ham}, the
observable $G_1(j,l)$ of Eq.~\eqref{eq:G1} for $p>2$ is a complex-valued function.
To further analyze its structure, it is useful to consider the Fourier transform of the operators $\hat F_j$
[see Eq.~\eqref{eq:Fourier}], and study the anyonic momentum distribution function (AMDF):
\begin{equation}
  n_0(k) = \frac{1}{L} \sum_{j,l} e^{+i k (j-l)} \, G_1(j,l) .
\end{equation}
A first inspection of the numerical results plotted in Fig.~\ref{Fig:Z3:Z6:nk} evidences two distinctive features,
which have been already pointed out in other anyonic models~\cite{Santachiara_2008}:
{\it i}) the absence of symmetry $k\to -k$, due to the lack of inversion symmetry of the model;
{\it ii)} the presence of a spike at $k_{\rm max}>0$. 
Following different arguments, we can qualitatively estimate the peak position to be located at
\begin{equation}
  k_{\rm max} = \kappa \pi N/L .
  \label{eq:kmax}
\end{equation}
First, in Sec.~\ref{Sec:FewBody} we have already observed that the anyonic gas behaves as a standard gas
with twisted boundary conditions, the twist of each momentum being $\kappa \pi/L$. As such, we can naively expect
that the AMDF is peaked around a wavevector equal to $N$ times such value.
Second, each time two FP operators are commuted, a phase $\omega$ appears. Assuming a gas with uniform density,
in order to compute the correlator $G_1(x,x+r)$, a number of $Nr/L$ anticommutations has to be performed,
and thus a phase $\omega^{Nr/L}$ is gained. This corresponds to a peak in the AMDF at $k_{\rm max}$.
Finally, according to Ref.~\cite{Calabrese_2007}, the low-energy theory predicts $G_1(x,x+r) \propto e^{i k_{\rm max}r}$,
where $k_{\rm max}$ is given by Eq.~\eqref{eq:kmax}.
The insets in both panels of Fig.~\ref{Fig:Z3:Z6:nk} show that such prediction works well only at low densities,
whereas for $N/L \gtrsim 1/4$ a significant discrepancy appears. We interpret this as a consequence of the fact
that in our model the local Hilbert space has a finite dimension $p$, whereas in the mentioned models it is infinite.

We conclude this part by mentioning that we have also numerically studied the anyonic correlation function
$G_2(x,y) = \langle \Psi_{\rm GS} | \hat F_x^{\dagger 2} \hat F_{x+r}^2 | \Psi_{\rm GS} \rangle$, obtaining similar results.
In particular, they display a power-law decay in qualitative agreement with the anyonic LL theory
of Refs.~\cite{Calabrese_2007, Bellazzini_2009}, although
larger discrepancies seem to emerge for $p=3$ (see App.~\ref{App:Anyon:CF} for further details).

\subsection{Fermionic correlation functions for $p=6$}\label{Sec:Fermi}
\label{subsec:fermi}

\begin{figure}
 \includegraphics[width=0.8\columnwidth]{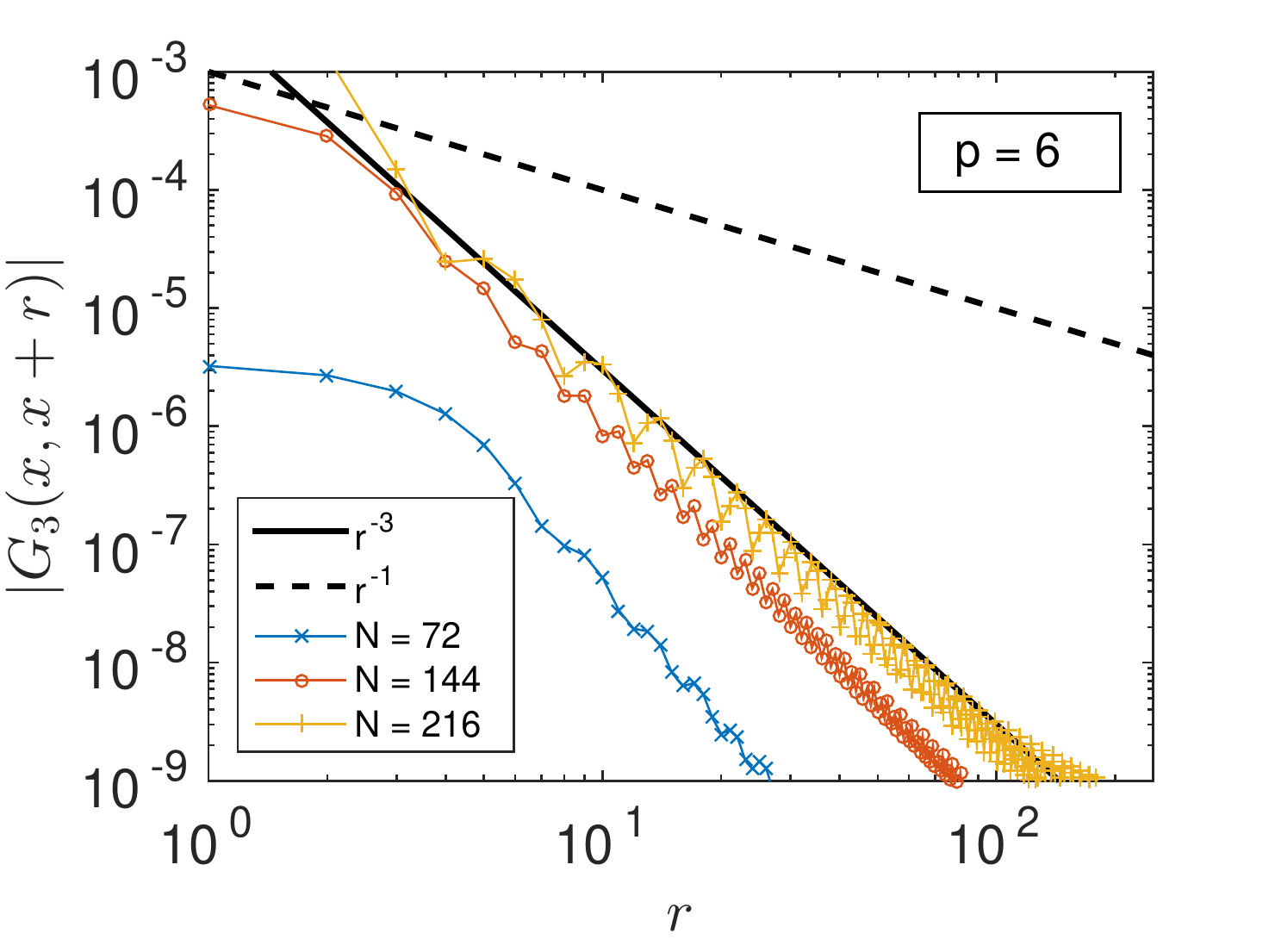}
 \includegraphics[width=0.8\columnwidth]{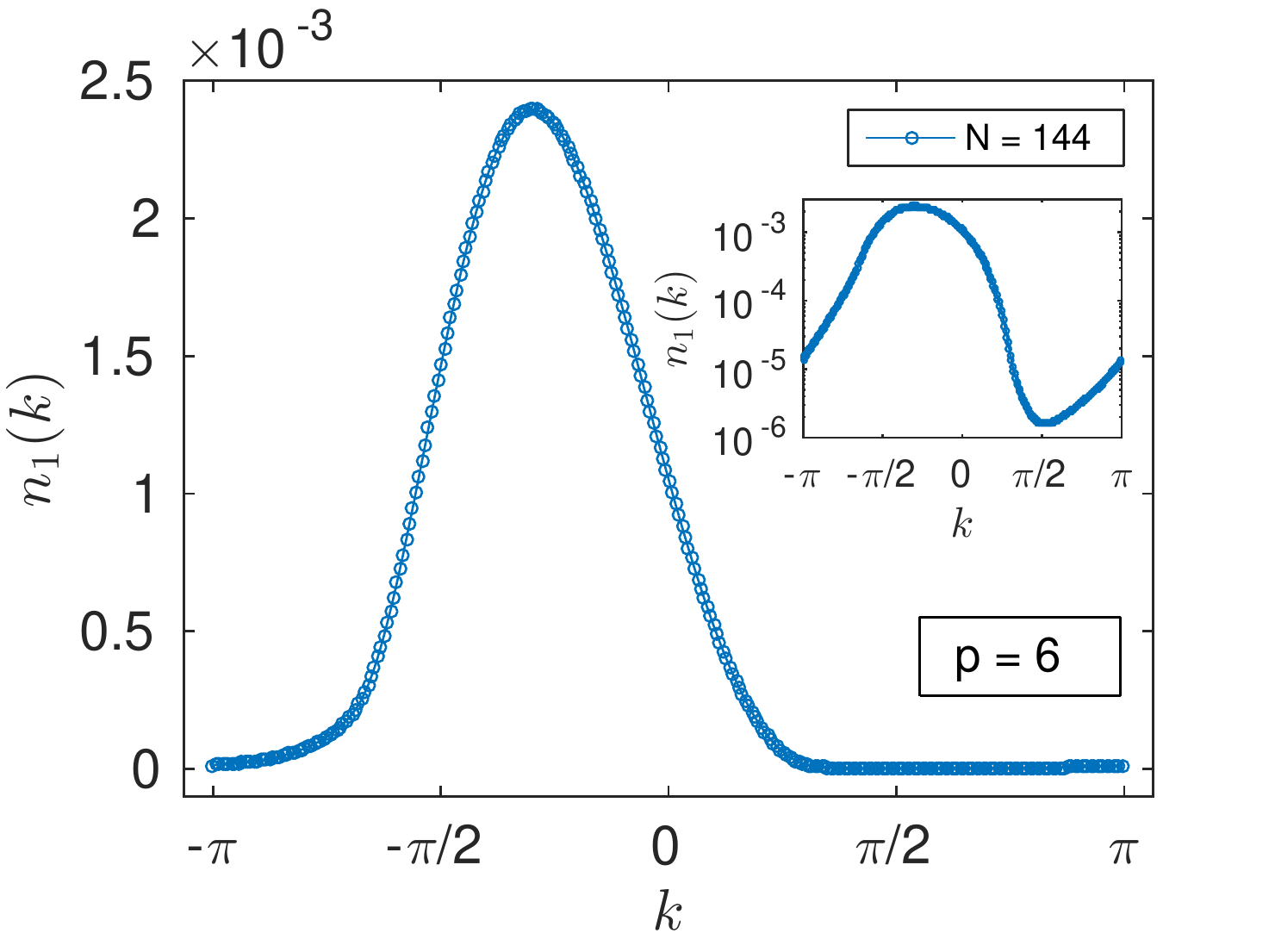}
 \caption{(Upper panel) Absolute value of the fermionic correlation function $G_3(x,x+r)$ as a function of $r$,
   for $p=6$ and several particle numbers. Data are for $L=288$. The bosonization prediction is superimposed
   as a thick black line, and that for free fermions as a dashed black line.
   Data points (symbols) have been fitted with a power-law function $|G_3| \sim r^{-\alpha_3}$, in the range $r \in [10, L/2]$.
   The resulting best-fit values of the exponent are: $\alpha_3 = 4.04 \pm 0.05$ ($N=72$),
   $\alpha_3 = 3.26 \pm 0.05$ ($N=144$), $\alpha_3 = 2.98 \pm 0.05$ ($N=216$).
   These can be matched with those for $\alpha_1$ (Fig.~\ref{Fig:Z3:Z6:G1}, lower panel):
   $9 \alpha_1 = 4.14 \pm 0,018$ ($N=72$), $9\alpha_1 = 3.46 \pm 0.02$ ($N=144$), $9\alpha_1 = 3.17 \pm 0.01$ ($N=216$).
   (Lower panel) Fermionic momentum distribution function $n_1(k)$ for $p=6$, $N=144$ and $L=288$.
   Inset: same plot in semi-logarithmic scale.}
 \label{Fig:Z6:G3:n1k}
\end{figure}

We now move to the study of fermionic operators $\hat f_j = \hat F_j^3$ introduced in Eq.~\eqref{Eq:Fermi:p6} for the case $p=6$.
We first define the fermionic correlation function
\begin{equation}
  G_3 (j,l) = \langle \Psi_{\rm GS} | \hat f_j^\dagger \hat f_l | \Psi_{\rm GS} \rangle.
\end{equation}
In Fig.~\ref{Fig:Z6:G3:n1k} we plot its absolute value and observe that it decays algebraically as $|G_3(x,x+r)| \sim r^{-{\alpha_3}}$. 
According to the approximate LL description~\cite{Calabrese_2007, Bellazzini_2009}, $\alpha_3 = 9 \alpha_1 = 18/p$.
In this specific case, $\alpha_3 = 3$. The fitted values are compatible with $9 \alpha_1$,
however the agreement increases with the density of the gas (see caption of Fig.~\ref{Fig:Z6:G3:n1k}).
At this stage a few remarks are in order. First notice that the exponent
\rev{of the decay rate of $G_3$} is clearly different from that of free fermions,
\rev{whose two-point correlation functions are known to decay as $r^{-1}$}.
Thus, a quadratic model of fractionalized fermions induces \rev{effective} strong correlations among \rev{quasiparticles}.
Moreover and importantly, for the larger density values, the prediction for the scaling of $|G_3(x,x+r)|$ quantitatively agrees
with the one predicted for a correlated state by Wen's hydrodynamics for a Laughlin state at filling $\nu=1/3$.

%%%%%%%%%%%%%%%%%%%%%%%%%%%%%%%%%%%%%%%%%%%%%%%%%%%%%%%%%%%%%%%%%%%%%%%%%%%%%%%%
%%%%%%%%%%%%%%%%%%%%%%%%%%%%%%%%%%%%%%%%%%%%%%%%%%%%%%%%%%%%%%%%%%%%%%%%%%%%%%%%

In the lower panel of Fig.~\ref{Fig:Z6:G3:n1k} we plot the fermionic momentum distribution function (FMDF):
\begin{equation}
  n_1(k) = \frac{1}{L} \sum_{j,l} e^{+ik (j-l)} \, G_3(j,l) .
\end{equation}
Although $n_1(k)$ does not exhibit the sharp discontinuity presented by the AMDF, it is again not $k \to -k$ invariant,
and has a maximum for a non-zero value of $k$. The form is roughly (but not exactly) symmetric around such point.
We stress that a similar FMDF cannot be easily traced back to any fermionic model, highlighting the impact of fractionalization.

To better rationalize this result, we can consider the scattering states introduced in Sec.~\ref{Sec:FewBody} for $p=4$
and the fermionic operators $\hat f_j = \hat F_j^2$.
It is interesting to observe that $\bra{\Psi} \hat f^\dagger_j \hat f_l \ket{\Psi} = |B|^2 e^{i (k+q) (l-j)}$.
The combination $k+q$ does not depend on the phase $\theta$ but depends on the statistical phase
through $k+q = \frac{2 \pi}{L} (\lambda_{k}+\lambda_q- 2/p)$ and thus the result is related to that of fermions
with twisted boundary conditions. For the ground state, $\lambda_k = \lambda_q =0$ and the peak of the FMDF
is shifted by $- \frac{2 \pi}{L} \frac{2} {p}$.

Before closing, we comment on a possible low-energy theory of the investigated lattice model.
Developing a microscopic bosonization theory of Fock parafermions starting from first principles is a task
that goes beyond the purposes of this article. However, based on the numerical observations collected for the gapless phase
for $p=6$, we {\color{black}can now} argue that the model shares some properties with a couple of counter-propagating Laughlin boundary modes
at filling factor $\nu=1/3$ with edge velocities $\pm v$. 

The latter represents an example of an anyonic Luttinger Liquid~\cite{Bellazzini_2009}, whose Hamiltonian can be written as
\begin{equation}
 \hat H = \frac{v}{2} \int \left[
(\partial_x \hat \theta)^2 + (\partial_x \hat \phi)^2
 \right]  {\rm d}x .
 \label{Eq:ALL}
\end{equation}
where $\hat \phi(x)$ and $\hat \theta(x)$ are the so-called dual fields and satisfy
$[\hat \phi(x_1), \hat \theta(x_2)] = i \frac{2\pi}p \Theta_{\rm H}(x_2-x_1)$,
and $\Theta_{\rm H}(x)$ is the Heaviside step function.
We can define low-energy right- and left-moving anyonic excitations using the bosonic fields
of Hamiltonian in Eq.~(\ref{Eq:ALL}), using the operators $\hat F_R(x)$ and $\hat F_L (x)$,
where $\hat F_{R/L} (x) \propto e^{i \alpha_{R/L}(\kappa) [ \hat \theta(x) \mp \hat \phi(x) ]}$.
Right- and left- movers are described by opposite statistical parameter, contained in the coefficient $\alpha_{R/L}(\kappa)$.
{\color{black}This property makes the anyonic Luttinger liquid time-reversal invariant.}

{\color{black}Also in the studied 1D lattice system we have both right and left movers, but the statistical parameter is unique,
  and the model is not time-reversal invariant.}
This motivates further investigation to establish the possible link between anyonic LLs and our Hamiltonian~\eqref{Eq:Ham},
where there is only one statistical parameter. The study of boundaries between two fractional quantum Hall states
closely separated by an insulating region started to attract significant attention in recent years~\cite{Lindner_2012, Clarke_2013}. 
In the end, our lattice model is well suited for developing a description for some of such boundaries that goes beyond
effective field theories with linearized dispersion relations.

\section{Conclusions}\label{Sec:Conc}

Motivated by recent proposals for an experimental realization of one-dimensional parafermionic systems in condensed-matter devices,
we addressed the simplest model of Fock parafermions, namely a tight-binding Hamiltonian.
The model is quadratic, but differently from its bosonic and fermionic counterpart, it does not enjoy an analytical solution. 
Our study exploits numerical methods and shows a number of remarkable properties that can be directly ascribed
to the exotic quantum statistics of parafermions, from the presence of bound states in the spectrum to the appearance of gapped phases.
Using arguments based on the level spacing statistics, we unambiguously demonstrate that the model is non integrable,
and rely on numerical methods for its characterization in the many-body case.
The remarkable feature of FPs is the fact that, in some cases, clusters of FPs behave as fractionalized fermions.
We show that, for $p=6$, our tight-binding Hamiltonian displays analogies with the low-energy properties of the boundary
between two neighboring Laughlin states, where fractionalized electrons counterpropagate.
This paves the way to test, in a lattice model, predictions that so far have only been checked in continuum field theories.
Moreover, it allows for a proper modelling of phenomena that require a beyond-LL description, including for instance curvature effects.
Finally, it has been highlighted that coupling two Hall bars with a Laughlin state each by alternating superconducting
and magnetic materials, it is possible to localize zero-energy parafermionic modes. 
Testing this prediction in a lattice tight-binding model with electronic superconductivity will be one of the next research directions.

\acknowledgments

We thank M.~Burrello, A.~Calzona, J.~De~Nardis, F.~Iemini, M.~Mintchev, C.~Mora, and R.~Santachiara for enlightening discussions on the subject.
This work was granted access to the HPC resources of MesoPSL financed
by the Region Ile de France and the project Equip@Meso (reference
ANR-10-EQPX-29-01) of the programme Investissements d’Avenir supervised
by the Agence Nationale pour la Recherche. M.~C.~acknowledges support from the Quant-Era project ``SuperTop''. M.~C.~S.~acknowledges support from the Israel Science Foundation, grants No.~231/14 and~1452/14.

\appendix

\section{Two-body scattering states}
\label{App:2body}

Let us consider the ansatz of Eq.~\eqref{Eq:2bodyAnsatz}.
From the projected eigenvalue equation $\bra{m_1,m_2} \hat H \ket{\Psi} = E \langle{m_1,m_2} \ket{\Psi}$ we can obtain three different equations:
\begin{widetext}
\begin{subequations}
\begin{align}
 E a(m_1,m_2) = -t \left[ 
 a(m_1-1,m_2)+a(m_1+1,m_2)+a(m_1,m_2-1)+a(m_1,m_2+1)
 \right], \qquad& m_2 > m_1+1, \label{Eq:BA:1}\\
 E a(m_1,m_2) = -t \left[ 
 a(m_1-1,m_2)+a(m_1+1,m_2)+\omega^*a(m_1,m_2-1)+a(m_1,m_2+1)
 \right], \qquad& m_2 = m_1+1, \label{Eq:BA:2}\\
 E a(m_1,m_2) = -t \left[ 
 a(m_1-1,m_2)+\omega a(m_1,m_2+1)
 \right], \qquad& m_2 = m_1. \label{Eq:BA:3}
\end{align}
\end{subequations}
Equation~\eqref{Eq:BA:1} admits a solution with $E(k,q) = -2t\cos(k) -2t \cos(q)$ for arbitrary values of $A$, $A'$, $B$, $k$ and $q$. 
Equations~\eqref{Eq:BA:2} and~\eqref{Eq:BA:3}
yield the following expression for $B/A$ and $e^{i \theta}$:
\begin{subequations}
\label{Eq:PhaseApp:Gen}
 \begin{align}
 \frac{B}{A} =& - \frac{e^{-ik}+e^{i \theta} e^{-iq} + \omega e^{iq} + \omega e^{i \theta} e^{ik}}{2[\cos(k)+ \cos(q)]}, \\
 e^{i \theta} =& - \frac{E(k,q)^2 e^{iq} +E(k,q) e^{i (q-k)} + E(k,q) e^{2iq} - \mathcal F(k,q)e^{-ik} - \omega \mathcal F(k,q) e^{iq}}{E(k,q)^2 e^{ik} +E(k,q) e^{i (k-q)} + E(k,q) e^{2ik} - \mathcal F(k,q)e^{-iq} - \omega \mathcal F(k,q) e^{ik}}, \qquad \mathcal F(k,q) = e^{i(k+q)}+ \omega^*,
 \label{Eq:PhaseApp}
 \end{align}
\end{subequations}
\end{widetext}
Although the above expressions are quite involved and it is not apparent, an explicit inspection of Eq.~\eqref{Eq:PhaseApp} shows that it is indeed a phase.
The numerical solution of Eqs.~\eqref{Eq:PhaseApp:Gen} allows for the determination of the wavevectors $k$ and $q$, and thus of the energy-momentum relation $E(k,q)$.

\section{Details on the LSS}
\label{App:LSS_Fermi}

As discussed in Sec.~\ref{Sec:LSS}, the tight-binding model of FPs exhibits level repulsion,
a fact that witnesses its absence of integrability.
Here we give further details on this issue.

\begin{figure}[!t]
  \includegraphics[width=0.9\columnwidth]{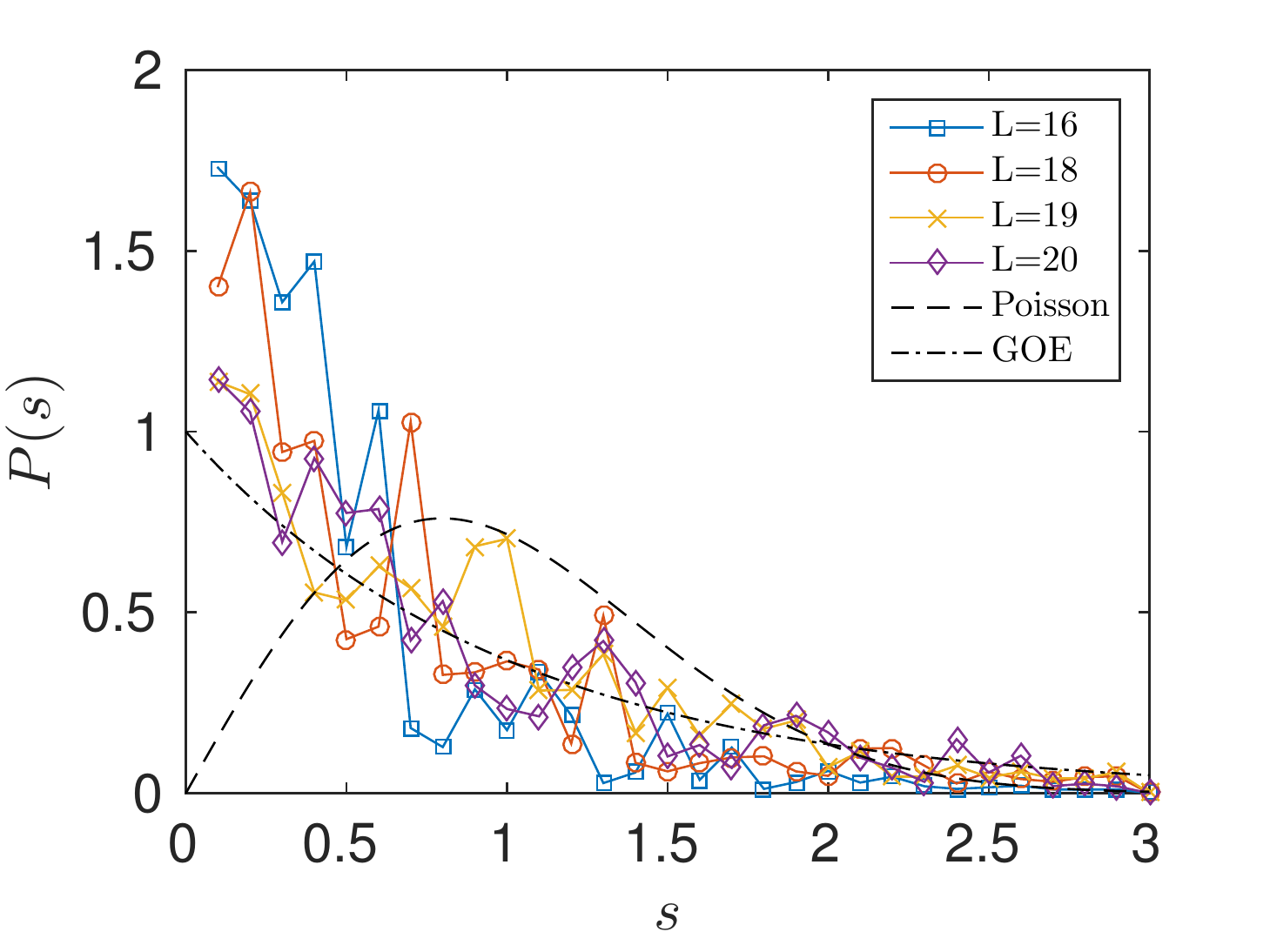}
  \caption{LSS for the fermionic tight-binding model [Eq.~\eqref{Eq:Ham} with $p=2$] and $N=7$ particles.
    We used OBC, added an inhomogeneity of strength $\varepsilon=10^{-2}$ on the first hopping,
    and a local chemical potential term of strength $\mu_1 = 10^{-2}$ on the first site.
    The various data sets are for different system sizes.
    The dashed curve corresponds to a GOE, while the dotted-dashed one to a Poissonian statistics.}
  \label{fig:LSS_Fermi}
\end{figure}

First of all, we explicitly show that, in the specific case of ordinary free fermions,
the situation is drastically different, the model being trivially integrable. 
We have computed the LSS for the Hamiltonian in Eq.~\eqref{Eq:Ham} with $p=2$,
adopting the same strategy employed for FPs aimed at breaking any obvious symmetry in the model,
such as translational invariance and inversion symmetry.
In particular, we have diagonalized the tight-binding Hamiltonian with OBC and for a fixed number of fermions.
To ensure that no trivial symmetries (as the inversion symmetry) are left,
we also admit an inhomogeneous hopping amplitude $t \to t + \varepsilon$ between the first two sites,
and a local chemical potential term of the form $-\mu_1 \hat F^\dagger_1 \hat F_1$.
The outcome of our ED simulations is presented in Fig.~\ref{fig:LSS_Fermi},
where the LSS is shown to converge to a Poissonian-like distribution, when increasing
the system size $L$.
In particular, notice the absence of level repulsion at small values of $s$
(typical of the WD surmise), which was shown to naturally emerge for models with $p>2$.

%%%%%%%%%%%%%%%%%%%%%%%%%%%%%%%%%%%%%%%%%%%%%%%%%%%%%%%%%%%%%%%%%%%%%%%%%%%%%%%%
%%%%%%%%%%%%%%%%%%%%%%%%%%%%%%%%%%%%%%%%%%%%%%%%%%%%%%%%%%%%%%%%%%%%%%%%%%%%%%%%

\begin{figure}[!t]
  \includegraphics[width=0.9\columnwidth]{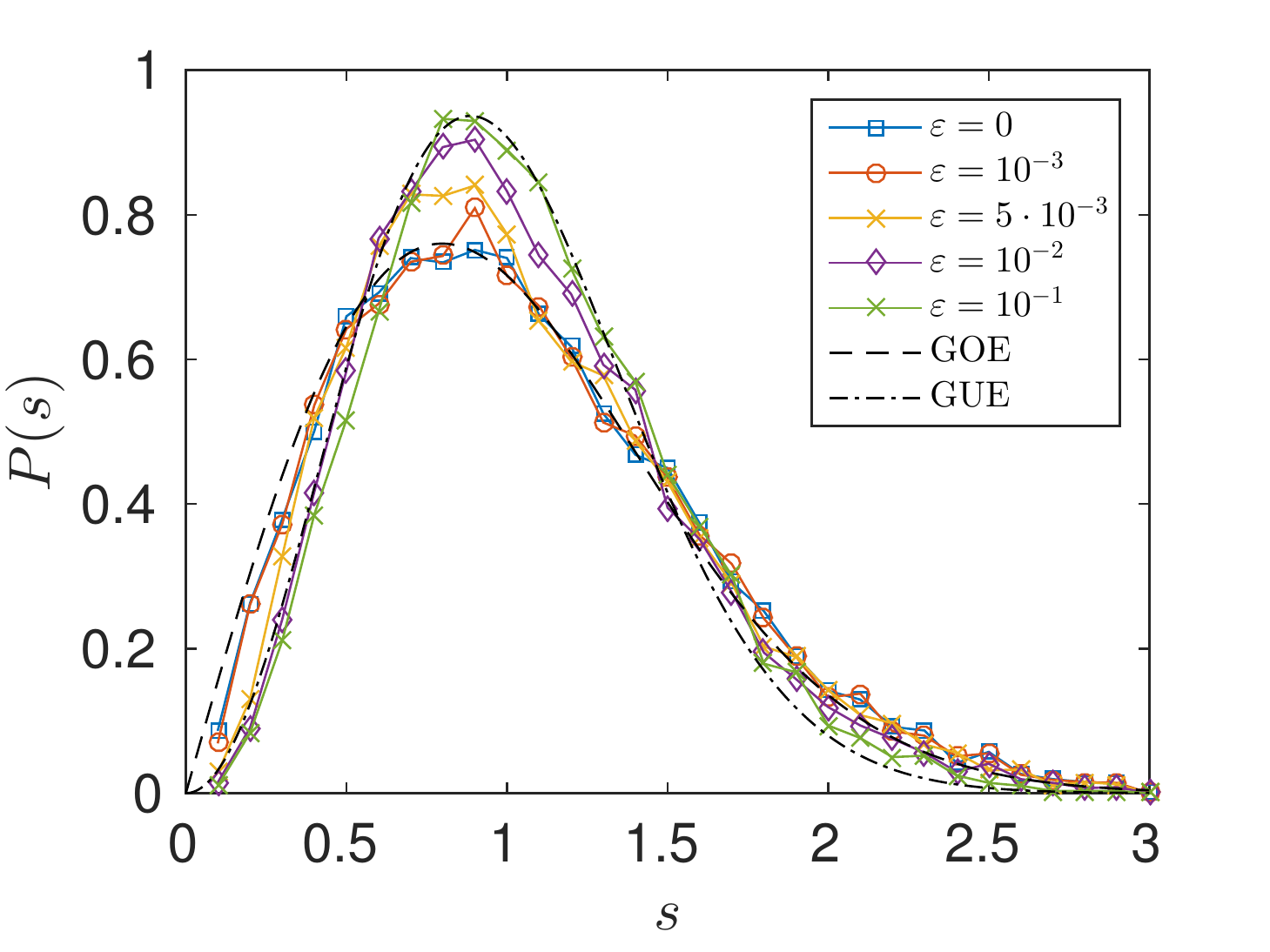}
  \caption{LSS for the tight-binding model of Eq.~\eqref{Eq:Ham}
    with $p=6$, $L=12$ sites, and $N=7$ particles. OBC have been enforced.
    The various data sets correspond to different values of the inhomogeneity $\epsilon$
    on the first hopping.
    The dashed curve corresponds to a GOE, while the dotted-dashed one to a GUE statistics.}
  \label{fig:LSSepsvar}
\end{figure}

Secondly, we have verified that the asymptotic WD distribution to which the LSS
of the FP spectrum converges is expected to depend on the specific symmetries of $\hat H$.
In fact, the numerical results presented in Sec.~\ref{Sec:LSS} support evidence that
our FP model (for $p>2$) obeys a WD statistics of the GOE type.
The latter is typical for systems which preserve an anti-unitary symmetry,
such as time-reversal. 
If we now consider a slightly different tunneling strength $t \to t + \varepsilon$
for the hopping term between the first and the second site,
Fig.~\ref{fig:LSSepsvar} shows that the shape of the resulting LSS exhibits
a crossover from GOE to GUE (as is typical for generic complex Hamiltonians).
We conclude by noticing that a rigorous analysis of the connection between
the Hamiltonian symmetries and the corresponding WD surmise for its LSS
is generally not obvious (see, e.g., Ref~\cite{Beenakker_1997})
and lies outside the purpose of the present study.

%%%%%%%%%%%%%%%%%%%%%%%%%%%%%%%%%%%%%%%%%%%%%%%%%%%%%%%%%%%%%%%%%%%%%%%%%%%%%%%%
%%%%%%%%%%%%%%%%%%%%%%%%%%%%%%%%%%%%%%%%%%%%%%%%%%%%%%%%%%%%%%%%%%%%%%%%%%%%%%%%

\section{Fradkin-Kadanoff transformation} \label{App:FKtrasf}
  In order to perform DMRG simulations, it is more convenient to preliminarily rewrite our model
in terms of conventional commuting operators, rather than using FPs which
obey the complicated anyonic commutation relations~\eqref{Eq:FP:CR2}. 
This can be done by means of a generalized Jordan-Wigner transformation
(also called Fradkin-Kadanoff transformation~\cite{Fradkin_1980}), which maps
the parafermions $\hat F_j^{(\dagger)}$ $(j=1,\ldots,L)$
to the Weyl hard-core boson matrices $\hat B_j^{(\dagger)}$ $(j=1,\ldots,L)$, according to:
\begin{equation}
  \hat F_j = \Bigg[ \prod_{k=1}^{j-1} \hat U_k \Bigg] \, \hat B_j \, .
  \label{eq:FK}
\end{equation}
The (now commuting) operators $\hat B_j$ and $\hat U_j$ have the following representations
in the Fock basis $\{ |m_j \rangle \}_{j=0}^{p-1}$:
\begin{equation}
  \hat B_j = \left(\begin{matrix}
    0 & 1 & 0 & \cdots & 0 \\
    0 & 0 & 1 & \cdots & 0 \\
    0 & 0 & 0 & \cdots & 0 \\
    \vdots & \vdots & \vdots & & \vdots \\
    0 & 0 & 0 & \cdots & 1 \\
    0 & 0 & 0 & \cdots & 0 
  \end{matrix} \right),
  \quad 
  \hat U_j = \left(\begin{matrix}
    1 & 0 & 0 & \cdots & 0 \\
    0 & \omega & 0 & \cdots & 0 \\
    0 & 0 & \omega^2 & \cdots & 0 \\
    \vdots & \vdots & \vdots & & \vdots \\
    0 & 0 & 0 & \cdots & 0 \\
    0 & 0 & 0 & \cdots & \omega^{p-1} 
  \end{matrix} \right),
\end{equation}
where $\omega=e^{2 \pi i/p}$. Notice that the onsite matrix representation for $\hat B_j$
is formally the same as the one for $\hat F_j$ [see Eq.~\eqref{Eq:Matrix:Elements}],
while $\hat U_j$ is a diagonal unitary operator with complex entries.

%%%%%%%%%%%%%%%%%%%%%%%%%%%%%%%%%%%%%%%%%%%%%%%%%%%%%%%%%%%%%%%%%%%%%%%%%%%%%%%%
\begin{figure}[t]
  \includegraphics[width=0.8\columnwidth]{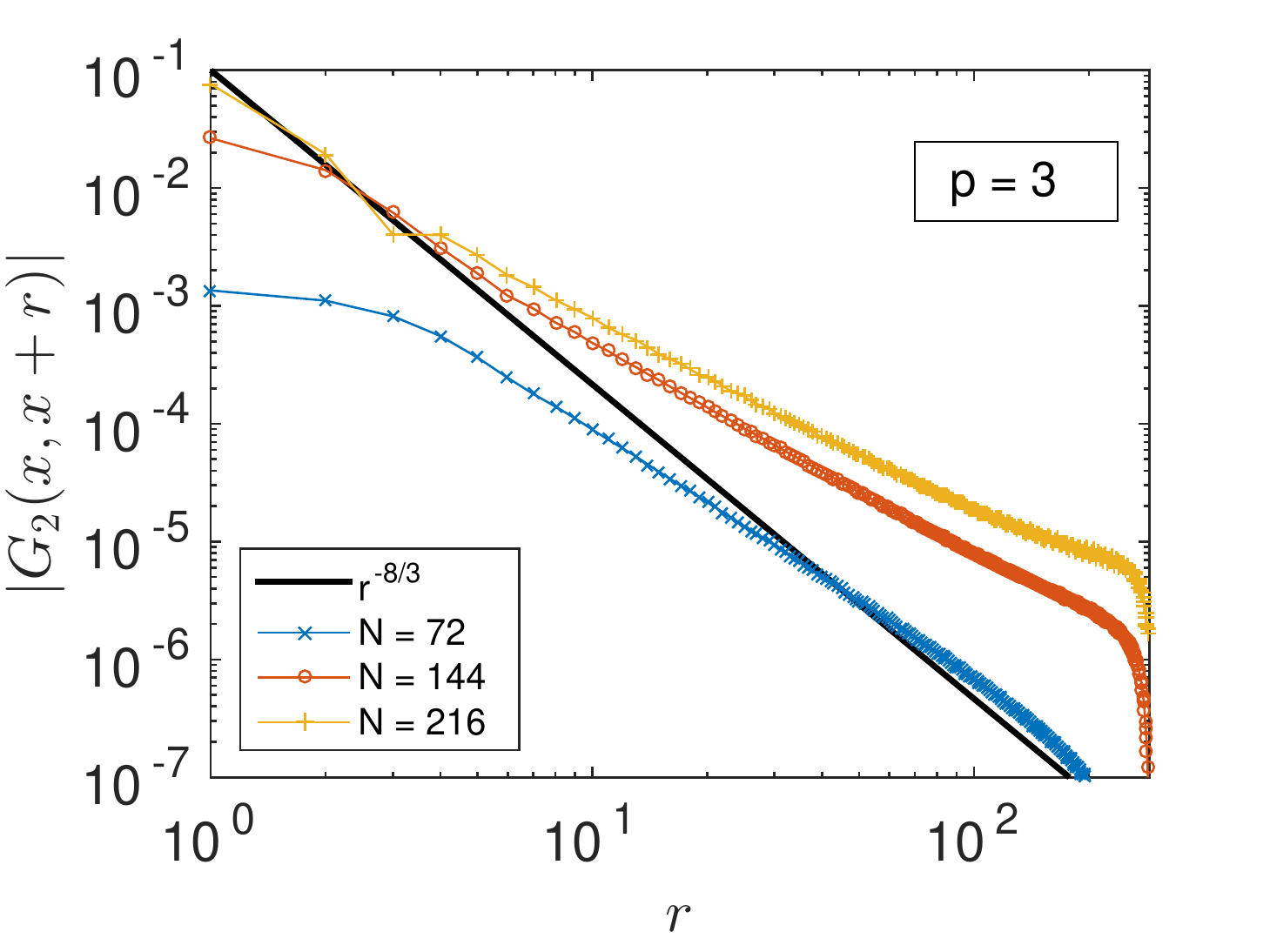}
  \includegraphics[width=0.8\columnwidth]{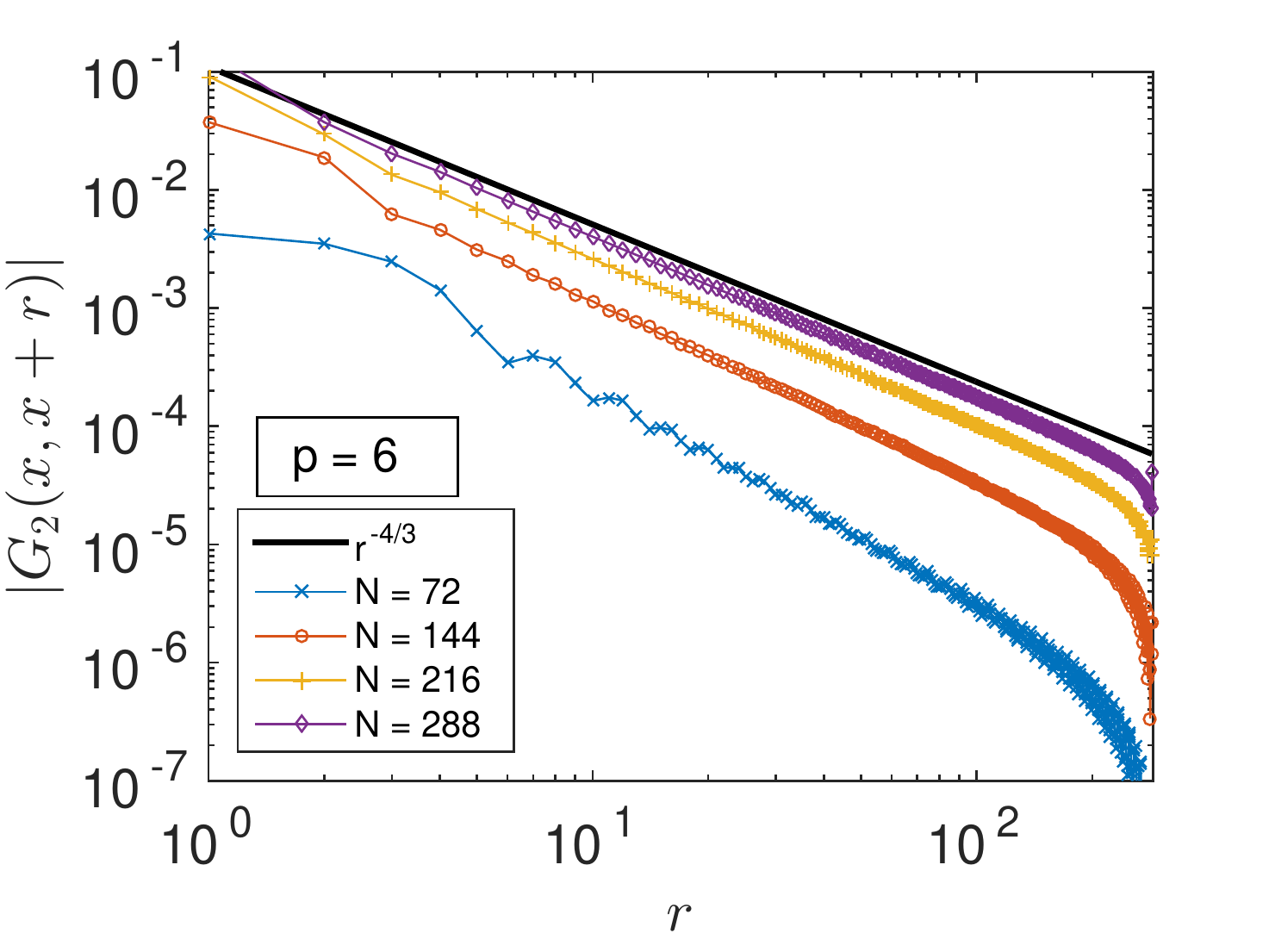}
  \caption{Absolute value of the anyonic correlation function $G_2(x,x+r)$ as a function of $r$,
    in the cases $p=3$ (upper panel) and $p=6$ (lower panel), for several particle numbers
    and a fixed chain length $L=288$.
    The LL prediction $r^{-8/p}$ is superimposed as a thick black line.
    For $p=6$, we have fitted the points $r \in [10, L/2]$ with a power-law function $|G_2| \sim r^{-\alpha_2}$,
    obtaining as best-fit parameter: $\alpha_2 = 1.85 \pm 0.03$ ($N=72$), $\alpha_2 = 1.55 \pm 0.02$ ($N=144$),
    $\alpha_2 = 1.42 \pm 0.01$ ($N=216$).}
 \label{Fig:Z6:G2}
\end{figure}
%%%%%%%%%%%%%%%%%%%%%%%%%%%%%%%%%%%%%%%%%%%%%%%%%%%%%%%%%%%%%%%%%%%%%%%%%%%%%%%%

Using Eq.~\eqref{eq:FK}, it is thus immediate to see that
\begin{align}
  \nonumber
  \hat F^\dagger_j \hat F_{j+1} \! & = \! \hat B^\dagger_j \big( \hat U^\dagger_1 \cdots \hat U^\dagger_{j-1} \big) \!
  \big( \hat U_1 \cdots \hat U_j \big) \hat B_{j+1} \! = \! \hat B^\dagger_j \hat U_j \hat B_{j+1}\\
  \nonumber
 \hat F^\dagger_{j+1} \hat F_j \! & = \! \hat B^\dagger_{j+1} \big( \hat U^\dagger_1 \cdots \hat U^\dagger_{j} \big) \!
  \big( \hat U_1 \cdots \hat U_{j-1} \big) \hat B_{j} \! = \! \hat B^\dagger_{j+1} \hat U^\dag_j \hat B_j
\end{align}
since the matrices $\hat U_k$ commute on different sites.
Therefore, the tight-binding FP Hamiltonian~\eqref{Eq:Ham} can be written in terms of
more manageable bosonic operators as:
\begin{equation}
  \hat H = -t \sum_j \left[ \hat B_j^\dagger \hat U_j \hat B_{j+1} + \hat U^\dagger_j \hat B_j \hat B_{j+1}^\dagger \right].
\end{equation}

We stress that, while the FP number operator $\hat N_j$ in Eq.~\eqref{eq:Nop}
maintains its usual representation in the bosonic language: $\hat N_j = \sum_l \hat B^{\dagger l}_j \, \hat B^l_j$,
the anyonic correlation functions are transformed into bosonic string correlators.
For example the $G_1$ function in Eq.~\eqref{eq:G1} becomes (for $j<l$):
\begin{equation}
  G_1(j,l) = \langle \Psi_{\rm GS} | \hat B_j^\dagger \big( \hat U_j \cdots \hat U_{l-1} \big) \hat B_l | \Psi_{\rm GS} \rangle.
\end{equation}

\section{Anyonic $G_2$ correlation functions}\label{App:Anyon:CF}

Here we discuss the results of our numerical simulations for the anyonic correlation function
\begin{equation}
  G_2(j,l) = \langle \Psi_{\rm GS} | \hat F_j^{\dagger 2} \, \hat F_l^2 | \Psi_{\rm GS} \rangle ,
\end{equation}
whose absolute value is reported in Fig.~\ref{Fig:Z6:G2}.
In analogy with the one-body density matrix $G_1$ reported in Eq.~\eqref{eq:G1} (see Sec.~\ref{subsec:gapless}),
we still observe a power-law decay of the type: $G_2(x,x+r) \sim r^{-\alpha_2}$. 
In this case, the LL theory~\cite{Calabrese_2007, Bellazzini_2009} predicts an exponent $\alpha_2 = 4 \alpha_1 = 8/p$,  
which nicely agrees with our data for $p=6$ (lower panel).
On the contrary, for $p=3$ the fitted power-law decay rates present some discrepancies
from the LL prediction (upper panel).
We have also evaluated the anyonic correlation functions $G_1(x,x+r)$ and $G_2(x,x+r)$
for $p=4$ and $p=5$ as well [not shown here], where LL relations for their power-law decay
are in accordance with our numerics. As such, we can ascribe the violation for the $G_2$ correlations
with $p=3$ to a truncation effect due to the dimensionality of the local Hilbert space.

\end{document}